\documentclass[paper,showpacs,preprintnumbers,amsmath,amssymb]{revtex4}
\usepackage{graphicx}

\begin{document}
\title{\bf Cosmological QCD phase transition in steady non-equilibrium dissipative Ho\v{r}ava-Lifshitz early universe
  }
 \author{M. Khodadi}%
\email{email:M.Khodadi@sbu.ac.ir}
\author{H. R. Sepangi}
\email{email:hr-sepangi@sbu.ac.ir}
\affiliation{%
Department of Physics, Shahid Beheshti University G. C., Evin, Tehran
19839, Iran
}%
\date{\today}

\begin{abstract}

We study the phase transition from quark-gluon plasma to hadrons in the early universe
in the context of non-equilibrium thermodynamics. According to the standard model of cosmology, a phase transition associated with chiral symmetry breaking after the electro-weak transition has occurred when the universe was about $1-10\mu s$ old.  We focus attention on such a phase transition in the presence of a viscous relativistic cosmological background fluid in the framework of non-detailed balance  Ho\v{r}ava-Lifshitz cosmology within an effective model of QCD. We consider a flat Friedmann-Robertson-Walker  Universe filled with a non-causal and causal bulk viscous cosmological fluid respectively and investigate the effects of the running coupling constants of  Ho\v{r}ava-Lifshitz gravity, $\lambda$, on the evolution of the physical quantities relevant to a description of the early universe, namely, the temperature $T$, scale factor $a$, deceleration parameter $q$ and dimensionless ratio of the bulk viscosity coefficient to entropy density $\frac{\xi}{s}$. We assume that the bulk viscosity cosmological background fluid obeys the evolution equation of the steady truncated (Eckart) and full version of the Israel-Stewart fluid, respectively.

\end{abstract}
\maketitle

\section{Introduction}
Perfect fluids, in contrast to real fluids, are in equilibrium because  their dynamics is reversible and hence produce no entropy and no frictional type heating. A realistic study of the early universe requires the invocation  of real relativistic  fluids. Great progress in relativistic thermodynamics was made by Eckart \cite{1} in the $1940$'s by extending irreversible thermodynamics from Newtonian to relativistic fluids. Since Eckart only considered  first-order deviation from equilibrium, his theory is known as the first-order theory of relativistic fluids. Although Eckart was concerned with special relativity his results are extendable to general relativity. In an Eckart fluid,  non-equilibrium states are described by the local equilibrium variables and hence dissipative perturbations propagate with infinite velocity and temperature which is the non-causal feature of the theory and unacceptable \cite{2}. Also, it is worth mentioning that the problem of infinite velocity of propagation  existed in non-relativistic thermodynamics  until 1967 when Muller proposed quadratic terms for thermal flux and viscous stress and succeeded to resolve this problem \cite{3}.  Later on, Israel along with Stewart applied Muller's extension to  Eckart theory and obtained modified dissipative relations that are acceptable from the point of view of relativistic thermodynamics \cite{4}. Their results are known as the second-order theory of relativistic fluids. It is useful to mention that after Israel and Stewart, other authors \cite{5} confirmed that Eckart's fluid suffers from serious drawbacks concerning causality and stability. They also developed a relativistic second-order theory which is known as the transient irreversible thermodynamics. The simplest model of a dissipative fluid is that which assumes the existence of bulk viscosity only. Qualitatively, we can regarded bulk viscosity as the macroscopic expression of microscopic frictional effects that appear in mixtures.

Turning our attention to the universe, in the simplest of cosmological models, there is no way to study entropy producing processes except through bulk viscosity. Bulk viscosity arises any time a fluid expands rapidly and ceases to be in thermodynamic equilibrium. The bulk viscosity, therefore, is a value of the pressure required to restore equilibrium to a compressed form of expanding system. Generally, unlike Eckart fluid, Israel and Stewart fluid is causal and stable but  are based on the assumption of small deviation from equilibrium, so that the transport equation is linear in the bulk viscous pressure. Linear assumption confirmed by most cosmological situations may  exit in some processes which involve more deviations from equilibrium. For example, in \cite{6}, the authors develop a nonlinear generalization of the causal linear thermodynamics of bulk viscosity for viscous fluid inflation. The big bang theory of cosmology dictates that the universe is expanding and this means that if we consider the evolution of the universe backwards in time, we will see the universe becoming denser and hotter. During this process the universe has undergone a series of transformations which could have occurred as phase transitions or crossovers as follows:\noindent\\
$(i)$ The GUT phase transition at $T_{c1}\sim {\cal O} (10^{14}-10^{16})$ GeV, where the symmetry between the strong and electro-weak interactions is spontaneously broken.\noindent\\ $(ii)$ The electro-weak phase transition at $T_{c2}\sim {\cal O}(10^{2})$ GeV, was a transition from a symmetric high temperature phase with massless gauge bosons to the Higgs phase where the $SU(2)\times U(1)$ gauge symmetry is spontaneously broken and all the masses in the model are generated.\noindent\\
 $(iii)$ The chiral symmetry breaking and the confining QCD phase transition at $T_{c3}\sim {\cal O}(10^{-1})$ GeV. Therefore, at roughly $10^{-6}\mu s$ after the Big Bang, matter was in the state of a quark-gluon plasma (QGP). As the plasma cooled,  quarks and gluons formed hadronic matter with the same density  as that of nuclear matter, that is $\rho\sim10^{15}\frac{gr}{cm^{3}}$.
 Studies of this process show that the system was likely in equilibrium so that static methods of statistical physics can be applied. The possibility of a phase transition in the gas of quark-gluon bags was demonstrated for the first time in \cite{7}. However, the phase transition in QCD can be characterized by a truly singular behavior of the partition function leading to a first or second order phase transition and, it may also be considered as a crossover transition with rapid changes in some observables, strongly depending on the values of the quark masses. The investigation of  any phase transition in the early universe necessitates  an equation of state (EoS) which plays an important role in the hot, strongly interacting matter and in the hydrodynamic description of heavy ion collisions \cite{9}. The EoS is described by thermodynamic quantities such as  pressure $(p)$, energy density $(\rho)$ or entropy density $(S)$ as a function of  temperature.

General Relativity is not a renormalizable theory which means that in  the UV limit it breaks down and cannot be quantized  using conventional quantization techniques. However, it was shown in \cite{K}  that theories whose action includes invariant quadratics in curvature
are renormalizable. Not surprisingly,  this comes at a  high price since  such theories  contain ghost degrees of freedom. The UV behavior of quantum gravity is one of the appealing topics in theoretical physics (especially string theory) which has been studied for a long time.  A new class of UV complete theories of gravity was proposed in the recent past by Ho\v{r}ava \cite{10} which is based on a scalar field theory proposed by Lifshitz  to explain the quantum critical phenomena in condensed matter physics \cite{11}. In the so-called Ho\v{r}ava-Lifshitz (HL) theory there is an asymmetry between time and space, that is
\begin{align}
t\rightarrow\ell^{z}t,\quad x^{i}\rightarrow \ell x^{i},
\end{align}
where $z$ is the dynamical critical exponent. Here, one may remark that $z$ can be considered as a gauge for describing the scale asymmetry between time and space. Indeed, by adding higher order spatial derivatives without adding higher order time derivatives, the behavior of the graviton propagator in the UV limit can be modified. Therefore HL gravity for  $z>1$ becomes non-relativistic. Briefly, the  crucial idea is related to  the breakdown of the fundamental role of the local Lorentz invariance in high energy. In other words, local Lorentz invariance is an accidental symmetry which appears only at low energies. Breaking the isotropic scaling of space and time  improves the UV behavior of the theory without introducing ghosts. Anisotropic scaling with
a dynamical critical exponent $z\geq3$ in HL gravity leads to  renormalizability of the theory based on power counting and reduces to Einstein's general relativity in the IR region. It is worth remembering that breaking the isotropic scaling of space and time leads to the breakdown of diffeomorphism invariance and the creation of additional scalar degree of freedom, i.e. a spin-0 mode for graviton. This
mode is potentially dangerous and may cause strong coupling problems that prevent the recovery of general relativity in the IR limit \cite{C.D, Cai}.  This may be regarded as a weak aspect of  HL gravity \cite{12}. It is well known that there are two natural classes of functions in HL theory that can be defined on a
foliation and there exists a special class of functions, the so-call projectables,
 which take constant values on each leaf of the foliation. Foliations can be equipped with a Riemannian structure. In Horava theory of gravity the  Riemannian structure consists of three objects; the metric $g_{ij}$ , the shift variable $N_{i}$ and the Lapse field $ N$ where $N$ is a projectable function, that is $N = N(t)$. In analogy  to Lifshitz scalar field theory, one may impose  an additional symmetry to the theory known as the
 detailed Balance. Hence it requires that potential part of the  action $S_{V}$ should be derivable from a superpotential $W $, that is $W=E^{ij}G_{ijkl}E^{kl}$,  where $E_{ij}$ follows from a variational principle
\begin{align}
E_{ij}=\frac{1}{\sqrt{g}}\frac{\delta W[g_{kl}]}{\delta g_{ij}}
\end{align}
 for which $G_{ijkl}$ is the inverse of the De Witt metric.  In view of the potential part of the action including  numerous terms containing various powers and spatial derivatives of curvatures of the spatial metric, the principle of detailed balance has been taken into consideration to limit such terms. However, this principle itself creates  some  problems, namely, the existence of a parity violating term, the (bare) cosmological constant has the opposite sign and it has to be much larger than the observed value and the scalar mode does not satisfy a sixth order dispersion relation and is not power-counting renormalizable  \cite{P.M}.   There are also other problems which are less obvious;  the infrared behavior of the scalar mode is plagued by instabilities and strong coupling at unacceptably low energies \cite{C.D}. On the other hand,  abandoning the detailed balance leads to a more complicated form of the theory but brings about some benefits including the suppression of the parity violating terms and the power-counting renormalizability. At the same time the cosmological constant is controlled and  not restricted \cite{P.M,  C.D}. Recently, Sotiriou, Visser and Weinfurtner (SVW),  discarding the principle of detailed balance but preserving the projectability condition,  showed that the most general such HL theory can
be properly formulated with eight independent coupling constants \cite{P.M}.  Among these eight coupling constants, one is
related to the kinetic energy responsible for  the spin-0 scalar graviton and the other seven are all related to the breaking of Lorentz invariance, which are highly suppressed by the Planck scale in the IR limit \cite{R}.

This study is organized as follows: In Section II  we present a brief review on non-local equilibrium  relativistic fluids with an emphasis on the Eckart and Israel-Stewart fluids.  In Section III  after a short review of non-detailed balance version of the HL gravity, the Friedmann equation of the FRW universe in the presence of a bulk viscous cosmological background fluid is derived.  In Sections IV we present a short survey of first order phase transition based on phenomenological models such as the MIT bag  \cite{47, 55, 56}, both for QGP  and hadron phase and obtain the  EoS  in the absence of the chemical potential $\mu=0$. In Sections V, using the EoS given by the MIT bag model at high temperature and an expansion for  low temperature in the hadron phase, we  study the dynamical behavior of the non-detailed balance HL early universe filled with non-causal and causal fluids  during the quark-hadron first order phase transition respectively. Finally, conclusions are drawn in section VII.

\section{Short review of relativistic fluids}
Events at space times are close to a fictional equilibrium state at that event thermodynamically. Based on non-casual Eckart theory, the local equilibrium quantities are denoted by $n,\rho,p,S,T$, representing number density, energy density, pressure, entropy and temperature respectively. The thermodynamic state of a simple relativistic  is generally characterized by an energy-momentum tensor $T^{\alpha\beta}$, a particle flow vector $N^{\alpha}=nu^{\alpha}$ and an entropy flow vector $S^{\alpha}$. For a dissipative fluid, the particle four-vector will have the same form as that of a perfect fluid and satisfies  $\delta N^{\alpha}\Delta_{\alpha\beta}=0$ where $\Delta_{\alpha\beta}$ is defined by $\Delta_{\alpha\beta}\equiv g_{\alpha\beta}-u_{\alpha}u_{\beta}$. In general $\delta N^{\alpha}\Delta_{\alpha\beta}=\psi$ where $\psi$ is a particle source if $\psi>0$ or sink if $\psi<0$.  Thus $\psi=0$  means no particle source or sink. This corresponds to choosing an average four velocity for which there is no dissipative particle flow, known as the particle frame. In  other words in a particle frame no particle is created or annihilated. Another standard frame is  the energy frame which satisfies  $\delta T^{\alpha\beta}u_{\alpha}\Delta_{\beta\mu}=0$, corresponding to an average four velocity in which there is no dissipative energy source.
 The satisfaction of the above two relations corresponds to the conservation of particle number and energy in the fluid system.
 For a dissipative fluid in the particle frame, the number density, energy density, temperature and four velocity are the same as that for a perfect fluid while the pressure has a deviation from the local equilibrium pressure, $p_{eff}=p+\Pi$, i.e. a splitting into the equilibrium part $p$ plus the bulk viscous pressure contribution $\Pi$. There is no convincing evidence that only the pressure should admit corrections due to dissipative processes \cite{T}. Bulk viscosity allows us to describe a mixture of different species effectively as a single fluid, provided a hydrodynamic description is reasonable \cite{6}. In an expanding fluid, the dissipation due to $\Pi$ leads to a decrease in kinetic energy and therefore in pressure so that $\Pi\leq0$. With the assumption that the fluid is moving orthogonal to the homogeneous spatial hypersurfaces, the four-velocity $u^{\alpha}$ is equal to the unit normal to such hypersurfaces. The energy-momentum tensor can be decomposed in terms of $u^{\alpha}$ according to
\begin{align}\label{e2}
T_{\alpha\beta}=(\rho+p_{eff})u_{\alpha}u_{\beta}+p_{eff}g_{\alpha\beta}+q_{\alpha}u_{\beta}+q_{\beta}u_{\alpha}+\pi_{\alpha\beta},
\end{align}
 where $\alpha,\beta$ assume $0\cdots 3$, $\pi_{\alpha\beta}$ , $q_{\alpha}$ and $u^{\alpha}$ are the anisotropic stress, heat conduction vector and the local equilibrium four velocity which are measured by an observer moving with the fluid, respectively. The four velocity $u^{\alpha}$ in the particle frame is defined by $u^{\alpha}=\frac{N^{\alpha}}{\sqrt{N^{\beta}N_{\beta}}}$ \cite{S.R}. Here, similar to the standard (perfect) fluid, the conservation law of energy-momentum, that is $\nabla^{\alpha}T_{\alpha\beta}=0$, is assumed to hold. In the co-moving frame the energy momentum tensor has the components $T^{0}_{0}=\rho,T^{1}_{1}=T^{2}_{2}=T^{3}_{3}=-p_{eff}$. It is also generally assumed  that there exists a linear relationship between the bulk viscous pressure $\Pi$ and the expansion $\theta$ in the form $\Pi=-\xi\theta$, where $\xi$ is the bulk viscosity coefficient and $\theta\equiv\nabla_{\alpha}u^{\alpha}$ is the expansion scalar, which for the FRW metric is equal to $\theta=3H$. Thus the bulk viscous pressure is
\begin{align}\label{e3}
\Pi=-3H\xi.
\end{align}
Also, the anisotropic stress term $\pi_{\alpha\beta}$ and heat conduction vector $q_{\alpha}$ are defined as
\begin{align}\label{e4}
\pi_{\alpha\beta}=-2\eta\sigma_{\alpha\beta}, \quad q_{\alpha}=-\lambda( D_{\alpha}T+T\dot{u}_{\alpha}),
\end{align}
where $\sigma_{\alpha\beta}$ is the shear tensor, $\eta$ is the shear viscosity and $\lambda$ is the thermal conductivity. According to relativistic Eckart fluid  there exists an acceleration term $q_{\alpha}$ arising from the inertia of heat energy. It is very interesting that unlike its Newtonian formulation  $\vec{q}=\lambda\nabla T$,  even in the absence of a temperature gradient, the heat flux arising from accelerated matter is non-zero. Therefore, the energy-momentum conservation equation is written as
\begin{align}\label{e5}
\dot{\rho}+3H(\rho+p_{eff})+D^{\alpha}q_{\alpha}+2\dot{u}_{\alpha}q^{\alpha}+
\sigma_{\alpha\beta}\pi^{\alpha\beta}=0.
\end{align}
Here, $D^{\alpha}$ is the covariant spatial derivative and $D^{\alpha}q_{\alpha}=h_{\mu}^{\alpha}h_{\alpha}^{\nu}\nabla_{\nu}q^{\mu}$. Also, $h_{\alpha\beta}$ is the projection tensor  defined by $h_{\alpha\beta}=g_{\alpha\beta}+u_{\alpha}u_{\beta}$. Unlike the perfect fluid for which the entropy flow vector is defined as $S^{\alpha}=sN^{\alpha}$, in Eckart fluid there is an added  dissipative term
\begin{align}\label{e6}
S^{\alpha}=sN^{\alpha}+\frac{q^{\alpha}}{T}.
\end{align}
The covariant form of the second law of thermodynamics is $\nabla_{\alpha}S^{\alpha}\geq0$ and the divergence of
the entropy current is given by
\begin{align}\label{e7}
T\nabla_{\alpha}S^{\alpha}=\frac{\Pi^{2}}{\xi }+\frac{q^{\alpha}q_{\alpha}}{\lambda T}+\frac{\pi_{\alpha\beta}\pi^{\alpha\beta}}{2\eta }.
\end{align}

The non-negativity  of the covariant divergence of entropy is required by the second law of thermodynamics and hence $\zeta ,\lambda,\eta,\geq0$.
It is known that the Israel-Stewart theory is related to linear  terms added to the entropy flow vector from perfect fluid (\ref{e6}).  Therefore we have the  general algebraic form
 \begin{align}\label{e8}
S^{\alpha}=sN^{\alpha}+\frac{q^{\alpha}}{T}-\left(\beta_{0}\Pi^{2}+\beta_{1} q_{\nu}q^{\nu}+\beta_{2}\pi_{\nu k}\pi^{\nu k}\right)\frac{u^{\mu}}{2T}
+\frac{\alpha_{0}\Pi q^{\mu}}{T}+\frac{\alpha_{1}\pi^{\mu\nu}q^{\mu}}{T},
\end{align}
where $\beta_{0}$, $\beta_{1}$ and $\beta_{2}$ are thermodynamic coefficients for scalar, vector and tensor dissipative
contributions to the entropy density, respectively. Also $\alpha_{0}$ and $\alpha_{1}$ are thermodynamic viscous and heat
coupling coefficients. The evolution equation of the bulk viscous pressure then reads
\begin{align}\label{e9}
\tau\dot{\Pi}+\Pi=-3\xi H-\frac{\epsilon}{2}\tau\Pi\left(3H+\frac{\dot{\tau}}{\tau}-\frac{\dot{\xi}}{\xi}-\frac{\dot{T}}{T}\right),
\end{align}
where $\tau$ is the relaxation time. In equation (\ref{e9}), $\epsilon=0$ gives the so-called truncated Israel-Stewart theory
with the evolution equation given by
\begin{align}\label{e10}
\tau\dot{\Pi}+\Pi=-3\xi H,
\end{align}
while the full theory has $\epsilon=1$. Obviously for the steady state, $\dot{\Pi}=\dot{\tau}=\dot{\xi}=\dot{T}=0$, the causal truncated Israel-Stewart theory (\ref{e10}) reduces to non-causal Eckart fluid, equation  (\ref{e3}). Relations (\ref{e9}) and (\ref{e10}) are valid only on the assumption that the fluid is close to equilibrium i.e. $|\Pi| \ll p$. Finally, it is worth mentioning that the conservation equations related to particle and energy frames are valid in Israel-Stewart fluid.

\section{Non-detailed balance HL gravity in the presence of bulk viscosity cosmological fluid}
In this section we shall present a  short introduction to Horava-Lifshitz  gravity without
detailed balance but with the projectability condition. As was mentioned before, the HL theory is constructed on the basic assumption of anisotropic scaling between space and time
\begin{eqnarray}
t\rightarrow\ell^{3}t,\quad x^{i}\rightarrow\ell x.
\end{eqnarray}
In a theory with anisotropic scaling, time and space are fundamentally distinct. It is then natural to use the ADM formalism for which the metric is given by
\begin{eqnarray}
ds^2=-N^2 c^{2}dt^{2}+g_{ij}\left(dx^{i}+N^{i}dt\right)\left(dx^{j}+N^{j}dt\right),
\end{eqnarray}
where $g_{ij}$, $N$ and $ N_{i}$ are the spatial metric, lapse and shift functions respectively. These variables are dynamical variables which under the above scaling behave as
 \begin{eqnarray}
 N\rightarrow N,\quad g_{ij}\rightarrow g_{ij},\quad N_{i}\rightarrow\ell^{2}N_{i},\quad N^{i}\rightarrow\ell^{-2}N^{i}.
 \end{eqnarray}

Because the original HL theory is power-counting renormalizable, the dimensions of the components of the above metric are
\begin{eqnarray}\label{em1}
[dx^{i}] =1 \quad [dt] =-3 \quad [N]=[g_{ij}]=0 \quad [N^{i}]=-2 \quad [ds^{2}]=2,
\end{eqnarray}
where the projectability condition demands a homogeneous lapse function, $N=N(t)$. It is useful to write the HL action in terms of geometric objects
characteristics of the ADM slicing of space-time, the $3D$ covariant derivative $\nabla_{i}$,
the spatial curvature tensor $R_{ijkl}$, and the extrinsic curvature $K_{ij}$  \cite{S.E}.
The action is then decomposed into a kinetic, a potential and a matter part
\begin{eqnarray}
S_{HL}=S_{k}+S_{v}+S_{m}.
\end{eqnarray}
The kinetic term is naturally given by
\begin{eqnarray}\label{em2}
S_{k}=\int dtdx^{3}N\sqrt{g}\alpha [K_{ij}K^{ij}-\lambda K^{2}],
\end{eqnarray}
where $K_{ij}$ $(K=K^{i}\,_{i})$ is the extrinsic curvature of spatial slices, defined by
\begin{eqnarray}
K_{ij}=\frac{1}{2N}(\partial_{t}  g_{ij}-\nabla_{i} N_{j}-\nabla_{j}N_{i}),
\end{eqnarray}
with $\nabla$ being the covariant derivative on the spatial slice whose metric $g_{ij}$ is used to raise and lower indices.
Indeed, the extrinsic curvature tensor  measures how the spatial slices in the ADM decomposition of spacetime
curve with respect to external observers \cite{J.P.M}. Note that $[K_{ij}]=3$ and $[\frac{2}{\kappa^{2}}]=0$. Using (\ref{em1}), it is not difficult to see that the dimension of potential is $6$. We also have  $[dV]=-6$ where  $dV=dtdx^{3}N\sqrt{g}$.
Hence the potential part in the most general action in the absence of detailed balance condition containing
terms with dimensions $\leq 6$ is given by
\begin{eqnarray}
S_{v-SVW}&=&\int dtdx^{3}N\sqrt{g}\left[ g_{0}\zeta^{6}+g_{1}\zeta^{4}R+g_{2}\zeta^{2}R^{2}+g_{3}\zeta^{2}R_{ij}R^{ij}+g_{4}R^{3}+
g_{5}R\left(R_{ij}R^{ij}\right)\right.\nonumber\\
&+&\left. g_{6}R^{i}_{j}R^{j}_{k}R^{k}_{i}+g_{7}R\nabla^{2}R+g_{8}\nabla_{i}R_{jk}\nabla^{i} R^{jk}\right],
\end{eqnarray}
where all the coupling constants $g_{i}$ are dimensionless and $[\zeta]=1$, for further details see \cite{P.M.V}. The Ricci scalar and tensor are given by $R_{ij}=g^{kl}R_{ijkl}$ and $R=g^{ij}R_{ij}$. Also the spatial curvature tensor, $R_{ijkl}$ is defined as
\begin{eqnarray}
R_{ijkl}=W^{i}\,_{il,k}-W^{i}\,_{jk,l}+W^{m}\,_{jl}W^{i}\,_{km}-W^{n}\,_{jk}W^{i}\,_{lm}. \quad W^{i}\,_{jl}=1/2g^{im}(g_{jm,l}+g_{ml,j}-g_{jl,m}),
\end{eqnarray}
where $W^{i}{}_{jl}$ are the Christoffel symbols. In the IR limit the HL action will simplify to
\begin{eqnarray}\hspace{-0.5cm}
S\rightarrow S_{I}\simeq \int dt dx^{3}N\sqrt{g}\left[\alpha(K_{ij}K^{ij}-\lambda K^{2})-g_{1}\zeta^{4} R-g_{0}\zeta^{6}\right].
\end{eqnarray}
Definitions
\begin{eqnarray}\hspace{-0.3cm}
x^{0}=ct \quad \lambda=\alpha= c=1,\quad g_{1}=-1,
\end{eqnarray}
will reduce this action to that of the Einstein-Hilbert's
\begin{eqnarray}
S_{EH}&=&\frac{1}{16\pi G}\int dx^{4}\sqrt{\widetilde{g}}(\widetilde{R}_{4}-2\Lambda_{EH})\\ \nonumber
&=&\frac{1}{16\pi G}\int dt dx^{3} \sqrt{g} N[K_{ij}K^{ij}-K^{2}+(R-2\Lambda)],
\end{eqnarray}
where $\zeta^{2}\equiv\frac{1}{16\pi G}$ and $\Lambda=\frac{g_{0}\zeta^{2}}{2}$. So, the cosmological constant $\Lambda$ is determined
by the free parameter $g_{0}$, and observationally $g_{0}\sim10^{-123}$ \cite{P.M.V}. Finally
\begin{eqnarray}
 S_{m}=\int dtdx^{3}N\sqrt{g}L_{m},
\end{eqnarray}
where $L_{m}=L_{m}(N,N_{i},g_{ij},\phi)$ is the Lagrangian density of matter fields, denoted collectively by $\phi$.
Now, in order to focus on a cosmological framework we use the $FRW$ metric
\begin{eqnarray}\label{em3}
N=1,\quad g_{ij}=a^{2}(t)\gamma_{ij},\quad N^{i}=0,
\end{eqnarray}
with
\begin{eqnarray}\label{em4}
\gamma_{ij}dx^{i}dx^{j}=\frac{dr^{2}}{1-kr^{2}}+r^{2}d\Omega_{2}^{2},
\end{eqnarray}
so that the homogeneous and isotropic metric is  written as
\begin{eqnarray}\label{em5}
ds^{2}=-dt^{2}+a(t)^{2}\left[\frac{dr^{2}}{1-kr^{2}}+r^{2}(d\theta^{2}+\sin^{2}\theta\, d\phi^{2})\right],
\end{eqnarray}
where $k=-1,\,0,\,1$ represents an open, flat or closed universe, respectively. We also suppose that the HL universe is filled with a viscous fluid.  Let us now define the shear tensor by $\sigma_{\mu\nu}=\theta_{\mu\nu}-\frac{1}{3}h_{\mu\nu}\theta$ where $\theta_{\mu\nu}=h^{\alpha}_{\mu}h^{\beta}_{\nu}u_{\alpha;\beta}$ is the expansion tensor and $\theta$ is the expansion scalar whose value in the FRW metric (\ref{em5}) is $\theta=3H$ . Explicitly, if we use this result to get the shear tensor we will obtain $\sigma_{\mu\nu}=0$. Thus from relation (\ref{e4}) we find that the anisotropic stress is zero,  $\pi_{\mu\nu}=0$. In addition, to maintain conservation of energy, an average four velocity should be chosen in such a way as to avoid the formation of dissipative energy sources or sinks  in the cosmological fluid system. Such a vector could be defined as \cite{S.R} $$u^{\alpha}= \frac{u^{\alpha}T^{\alpha\beta}u_{\beta}}{\sqrt{T^{\mu\nu}u_{\nu}T_{\mu\gamma}u^{\gamma}}},$$
which, in conjunction with the definition of heat conduction vector (\ref{e4}), leads to $q_{\alpha}=0$. Therefore, the energy-momentum tensor of the viscous fluid (\ref{e2}) in the background of metric (\ref{em3}) will reduce to
\begin{align}\label{e14}
T_{\alpha\beta}=(\rho+p_{eff})u_{\alpha}u_{\beta}+p_{eff}g_{\alpha\beta},
\end{align}
where $\rho$, $p_{eff}$ are the energy density and isotropic effective pressure, respectively. Also  $u^{\alpha} = (u^{0}, u^{\mu})$ where $\mu= 1,2,3$. In comoving coordinates $u^{0} = 1$ and $u^{\mu} = 0$.  We thus obtain the gravitational field equations of the HL theory, corresponding to the line element (\ref{em3}) and (\ref{em4}) as
\begin{eqnarray}\label{e15}
3(3\lambda-1)H^{2}=6\kappa^{2}\rho+\beta_{1}+\frac{\beta_{2}k}{a^{2}}+\frac{\beta_{3}k^{2}}{a^{4}}+\frac{\beta_{4}k}{a^{6}},
\end{eqnarray}
\begin{eqnarray}\label{e16}
(3\lambda-1)(\dot{H}+H^{2})=-\kappa^{2}(\rho+3p_{eff})+\frac{\beta_{1}}{3}-\frac{\beta_{3}k^{2}}{3a^{4}}-
\frac{2\beta_{3}k}{3a^{6}},
\end{eqnarray}
where
\begin{eqnarray}
\kappa^{2}=\frac{1}{6\zeta^{2}},\quad \beta_{1}=\frac{g_{0}\zeta^{6}}{6},\quad \beta_{2}=-6g_{1}\zeta^{4}>0,\quad \beta_{3}=12\zeta^{2}(3g_{2}+g_{3}),\quad \beta_{4}=24(9g_{4}+3g_{5}+g_{6}).
\end{eqnarray}
As was noted above,  $g_{0}\ll1$ regardless of  $\beta_{1}$ and  $\lambda$ is a dimensionless running coupling constant in the  Friedmann equations.
Also the energy conservation law $\nabla_{\alpha}T^{\alpha\beta}=0$ gives the continuity equation in the form
\begin{align}\label{e17}
\dot{\rho}+3H(\rho+p_{eff})=0.
\end{align}
The existence of the terms $1/a^{2}$, $1/a^{4}$ and $1/a^{6}$ in the Friedmann equations (\ref{e15}) and (\ref{e16})  correspond to  spatial
curvature contributions, radiation pressure or dark radiation and stiff matter respectively.
It is worth remarking that the stiff-matter term does not appear in the detailed balance version of the HL theory. The potential for the detailed balance case is given by
\begin{eqnarray}
\pounds_{V-DB}=\beta C_{ij}C^{ij}+\gamma\epsilon^{ijk}R_{il}\nabla_{j}R^{l}_{k}+\varsigma R_{ij}R^{ij}+\zeta R^{2}+\eta R+\sigma,
\end{eqnarray}
where $C^{ij}$ is Cotton tensor defined as $$C^{ij}=\epsilon^{ikl}\nabla_{l}\left(R^{j}_{l}-\frac{1}{4}R\delta^{j}_{l}\right).$$
In the above potential only the term $C^{i}\,_{j}C^{j}\,_{i}$ has dimension $[C^{i}\,_{j}C^{j}\,_{i}]=6$. Beside, the Cotton tensor  disappears  for the
constant curvature spatial slices of a FRW spacetime. Therefore,  in the Friedmann equation of detailed balance HL gravity we remove the term $1/a^{6}$. Since the early universe is known to be flat to a good approximation, in what follows we  will focus attention on the non detailed balance condition with $k=0$ and cosmological constant $\Lambda=0$.

Finally,  it is worth mentioning that despite all the phenomenological difficulties  many efforts have been made to determine the coupling constants of the HL theory. For example,  in \cite{S.D} the authors have used  cosmological observations  to impose  constraints on the running coupling constants of HL cosmology and have reported that to within $1\sigma$, that the running coupling constant $\lambda$ is bounded very close to its IR value, $|\lambda-1|\lesssim0.02$.

\begin{figure}
\includegraphics[width=8 cm]{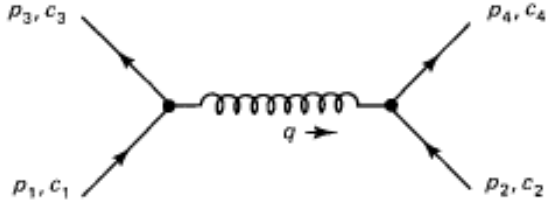}
\centering\caption{The quark-quark interaction to lowest order, figure from the first reference in \cite{D.H}.{\label{fig.F0} \small}}
\end{figure}

{\section{First order QGP-hadron phase transition}}
A phase transition is a phenomenon in which a drastic change between thermodynamic phases occurs as the system parameters such as energy density  $\rho$ or entropy density $S$ are varied.  In the early universe, the last phase transition predicted by the standard model of particle physics took place at the QCD scale $T \sim 200$ MeV  \cite{54}. When the universe was about $t \sim 10^{-6}-10^{-5}s$ old,  the Hubble radius was around 10 km corresponding to scales of about $1$ pc or about $3$ light years today. The quark-hadron phase transition is a fundamental notion to study  particle physics and is an essential part of any study dealing with the underlying mechanisms responsible for the evolving universe at its early stages of formation. The thermodynamics of the quark-gluon phase can be considered either in QCD  or using an ``effective interaction'' approach for which \cite{47} and \cite{55, 56} are two examples. Interestingly, both approaches agree on the same transition temperature but may disagree on the order of the transition, i.e. whether it is of first order, second order and or smooth crossover. In order to study the phenomena of quark-hadron phase transition in the framework of non-detailed balance HL  universe with a bulk viscosity cosmological background fluid, we have to specify  the EoS of the  QGP and hadron phase in the presence of the bulk viscous. Therefore, we should adopt an ``effective interaction'' approach. To this end,  we assume that quarks have different flavors, $u$ and $d$, in the QGP phase. To lowest order, figure \ref{fig.F0}, the amplitude of interaction $u+d\rightarrow u+d$ is given by
\begin{align}
\emph{M}_{qq}=-\frac{g_{s}^{2}}{4}\frac{1}{q^{2}}[\bar{u}(3)\gamma^{\mu}u(1)][\bar{u}(4)\gamma_{\mu}\bar{u}(1)]
(c^{\dag}_{3}\lambda^{\alpha}c_{1})(c^{\dag}_{4}\lambda^{\alpha}c_{2}),
\end{align}
 where $g_{s}=\sqrt{4\pi \alpha_{s}}$ is the well known strong coupling constant, $c_{i}$ are column vectors representing the color of the corresponding quarks,  $\lambda^{\alpha}$ $(\alpha=1, 2, ....8)$ are Gell-Mann matrices and $u, d$ are Dirac spinor-field operators representing the two light-quark fields.
We note that the above amplitude is similar to the electron-muon $e+\mu\rightarrow e+\mu$ scattering
\begin{align}
\emph{M}_{e\mu}=-\frac{g_{e}^{2}}{(p_{1}-p_{3})^{2}}[\bar{u}^{(s_{3})}(p_{3})\gamma^{\mu}
u^{(s_{1})}(p_{1})][\bar{u}^{(s_{4})}(p_{4})\gamma_{\mu}u^{(s_{2})}(p_{2})].
\end{align}
The only the difference is that $g_{e}=\sqrt{4\pi \alpha}$ should be replaced with $g_{s}$ and the color factor $C_{f}=(c^{\dag}_{3}\lambda^{\alpha}c_{1})(c^{\dag}_{4}\lambda^{\alpha}c_{2})$ is taken into account.  One would therefore deduce that inter-quark potential reads
\begin{align}
U_{qq}=C_{f}\frac{\alpha_{s}}{r},
\end{align}
where $C_{f}=\frac{N^{2}_{f}-1}{2N_{f}}$,  with $N_{f}$ being the number of quark flavors.  Therefore, one adds the interaction potential between quarks to the total energy of any particle in the ideal gas model, i.e., $E_{total}=\sqrt{p_{i}^{2}+m_{i}^{2}}+U_{qq}$.
An important property of QCD at high temperatures is the restoration of chiral symmetry. For two massless quarks the chiral symmetry group is $SU(2)_{L}\bigotimes SU(2)_{R}$, which is isomorphic to $O(4)$. At low temperatures the symmetry breaks spontaneously to $SU(2)_{L=R}$ which is isomorphic to $O(3)$. In a similar fashion as in condensed matter physics, the breakdown of  $O(4) \rightarrow O(3)$ symmetry points to a second order phase transition. One would therefore expect that with two massless quarks the QCD phase transition is of second order. However, for three massless quarks, one expects a first order phase transition, since systems with symmetry breaking $SU(3)_{L}\bigotimes SU(3)_{R}\rightarrow SU(3)_{L=R}$ do not have second order phase transitions. Of course, quarks are not completely massless; the s-quark has an intermediate mass and it is unclear if the
two or three flavor argument applies to our universe. No doubt a true second order phase transition would only happen in the limit of vanishing quark masses.
Hence, in our universe we either had a first order phase transition or merely a cross over.

Next we shall use the simplest possible form of an EoS incorporating a first-order phase transition, a bag EoS which has often been used for the zero-baryon-number and also any baryon-number cases. Since the ratio of baryon-entropy is very small in the early universe, we shall restrict ourselves to the case where the chemical potential is very small, $\mu \ll T$, meaning that we can determine the EoS in the absence of chemical potential. Both
theoretical considerations \cite{P.K} and experimental data collected from Relativistic Heavy Ion Collider (RHIC) suggest that the matter created in heavy ion collisions behaves like a strongly coupled liquid with very low viscosity. Hence, the MIT bag model that effectively includes strong interactions via a bag constant can be of use as an equation of state. The MIT bag model was one of the first phenomenological quark models that implemented two of the key properties of QCD,  asymptotic freedom and confinement, in a constitutive way. In the MIT bag model, hadrons are made up of a gas of free (or only weakly interacting) quarks and gluons which are confined to a finite region of space, the bag. We start from the equation of state of matter in the quark phase which can generally be given in the form
\begin{align}\label{e39}
\rho_{q}=3a_{q}T^{4}+V(T)+U_{qq},\quad p_{q}=a_{q}T^{4}-V(T)+U_{qq},
\end{align}
where $a_{q}=\frac{\pi^{2}}{90}g_{q}$ with $g_{q}=g_{Q}+(g_{b}+\frac{7}{8}g_{f})$ and $g_{Q}=16+(21/2)N_{f}$ is the intrinsic statistical weights for the quark-gluon plasma and $g_{b}, g_{f}$ are the statistical weights of bosons and fermions, respectively. At the epoch of the quark-hadron phase transition photons $(g_b=2)$, electrons $(g_f=4)$, muons $(g_f=4)$ and neutrinos $(g_{f}=6)$ yield $g_{b}+\frac{7}{8}g_{f} =14.25$. Here $N_{f}$ is two at lower temperature corresponding to the $u$, $d$ quarks and three at higher temperatures where the strange quark becomes relativistic, hence $ g_{q}=51.25$.

An effective Lagrangian relevant to the study of QGP-hadron phase transition admitting the chiral symmetry of QCD has been proposed in \cite{57} and is given by
 \begin{align}
\pounds=\sum_{k=1}^{n_{f}}\left[i\bar{\psi_{k}}\gamma^{\mu}\partial_{\mu}-g\zeta\left(\bar{\psi}^{L}_{k}U\bar{\psi}^{R}_{k}+
\bar{\psi}^{R}_{k}U^{\dagger}\bar{\psi}^{L}_{k}\right)\right]+\frac{1}{2}\partial_{\mu}\zeta\partial^{\mu}\zeta+
\frac{1}{2}\partial_{\mu}\pi\partial^{\mu}\pi+\frac{1}{4}\mbox{Tr}(\partial_{\mu}U\partial^{\mu}U^{\dagger})-
V(\zeta),
\end{align}
where $\zeta=\sqrt{\sigma^{2}+\pi^{2}}$, $U$ is an element of $SU(2)$ defined by $\zeta U=\sigma+i\tau.\pi$ and $\psi^{L,R}_{k}$  refers to  the left and right handed components of the quark field $\psi_{k}$. As can be seen from the above Lagrangian, quark fields $\psi_{k}$ are interacting with  $\zeta$ which is formed from the $\pi$ meson field and the scalar field $\sigma$. The quantity $\zeta$ is a well know  chiral field  since $(\pi^{2}+\sigma^{2})$ is invariant under vector and axial vector transformations \cite{V.K}.  Usually, the general form of a self-interacting potential $V(\zeta)$ includes a quartic function in terms of  $\zeta$.  Applying temperature effects at $\zeta=0$ we find, second reference in \cite{57}
\begin{align}\label{e40}
V(T)=B+\gamma_{T}T^{2}-\alpha_{T}T^{4}.
\end{align}
Note that $V$ at $\zeta=0$ has a local minimum corresponding to a chirally symmetric, quasi-stable non-physical vacuum with  energy density $B$ and at $\zeta=f_{\pi}=93MeV$ has an absolute minimum corresponding to a physical vacuum. In other words, breaking  chiral symmetry  corresponds to transition $\zeta=0\rightarrow\zeta=f_{\pi}$ in the self-interacting potential function.  For more detail the reader is referred to the appendix of third reference in \cite{57}. The quantity $B$ is the QCD vacuum energy or bag pressure constant. The bag is stabilized by a term of the form $g^{\mu\nu}B$ which is added to the energy-momentum tensor inside the bag, since the energy-momentum tensor of a perfect fluid in the rest frame is given by $T^{\mu\nu}=\mbox{diag}(\rho,p,p,p)$ and hence $B$ can be interpreted as a positive contribution to the energy density and a negative contribution to the pressure inside the bag. The MIT bag model cannot say anything about the origin of the non-trivial vacuum, but treats $B$ as a free parameter. Results obtained in low energy hadron spectroscopy, heavy ion collisions and phenomenological fits of light hadron properties give $B^{1/4}\in(100-200)MeV$, $\alpha_{T}=\frac{7\pi^{2}}{20}$ and $\gamma_{T}=\frac{m_{s}^{2}}{4}$  \cite{58}, where $m_{s}$ is the  the strange quark mass taken in the range $m_s \in(60 - 200)MeV$. Evolution of the running coupling constant of QCD is governed by the equation
\begin{align}
\alpha_{s}(q)=\frac{12\pi}{(33-2n_{f})ln(q^{2}/\Lambda^{2}_{QCD})},
\end{align}
where $n_{f}$ is the number of quark flavors and $\Lambda_{QCD}$ is a scale parameter of QCD. Under the condition $q^{2}\gg\Lambda^{2}_{QCD}$ the effective constant $\alpha_{s}(q)\ll 1$, that is, at short distances (high temperatures) QCD is a perturbative theory with asymptotic freedom. It is worth mentioning that for $q^{2}\leq\Lambda^{2}_{QCD}$ one cannot use  perturbation theory as strongly interacting gluons and quarks start to form bound states of hadrons. Hence predicted values of the free parameter $\Lambda_{QCD}$ is very important. The exact value of $\Lambda_{QCD}$ still remains unknown but its approximate value is in the range $100-200MeV$. Therefore, in line with most models to respect   asymptotic freedom in QGP phase we set $\alpha_{s}\approx 0$ and accordingly $U_{qq}\approx0$.  Needless to say that this assumption is not always valid; in quark stars near the normal nuclear density, quarks strongly interact with each other \cite{N}. Therefore the EoS, equation (\ref{e39}) reduces to
\begin{align}\label{e41}
\rho_{q}=3a_{q}T^{4}+V(T),\quad p_{q}=a_{q}T^{4}-V(T),
\end{align}

Usually thermodynamic quantities follow from  pressure. In lattice QCD however, one introduces a dimensionless quantity called the ``interaction measure,'' defined as $\frac{\rho-3p}{T^{4}}$. This is basically the trace of the energy momentum tensor divided by $T^{4}$  and  vanishes if the theory is conformally symmetric  \cite{R.D}.  In a system with conformal invariance the bulk viscosity vanishes. The  substitution of (\ref{e41}) in the ``interaction measure'' together with the self-interaction potential $ V(T)$ leads to  $\frac{\rho-3p}{T^{4}}\neq0$ with the consequence of  EoS for  QGP phase breaking conformal invariance with the existence of a finite bulk viscosity in the system. We expect however to have a negligible value at high temperatures \cite{P.A}. Generally, in any system where the interaction is turned off, conformal invariance makes the bulk viscosity to be zero. from hereon we assume interaction among  particles in order to find the Eos hadron phase when a bulk viscous fluid is present. 

The first approach for the pion gas is based on the Weinberg theory of pion interactions \cite{100} which uses a nonlinear Lagrangian containing all processes quadratic in pion momentum. In the first and third references in \cite{54}, corrections of first and second orders in the Weinberg Lagrangian were estimated and it has been shown that taking into account the pion re-scattering leads to nonzero corrections to the pressure and deviations from the ideal pion gas formula
\begin{align}\label{e4-1}
p_{\pi}\simeq\frac{\pi^{2}}{30}T^{4}\left[1+\left(\frac{T}{T_{int}}\right)^{4}\right],\quad T_{int}\simeq150MeV.
\end{align}
Using thermodynamics relation, $\rho=T\frac{dp}{dT}-p$, the energy density is given by
\begin{align}\label{e4-2}
\rho_{\pi}\simeq\frac{\pi^{2}}{10}T^{4}\left[1+7/3\left(\frac{T}{T_{int}}\right)^{4}\right].
\end{align}
In the absence of viscosity  the hadron phase is regarded by a cosmological fluid with energy density $\rho_{h}$ and pressure
$p_{h}$ proportional to $T^{4}$ as an ideal gas of massless pions \cite{F.M.H}. Obviously, the added term proportional to $T^{8}$ in the EoS, equations  (\ref{e4-1}) and (\ref{e4-2}), represents  self interacting massless pions. In other words,  in the presence of viscosity the assumption of ideal gas of massless pions  fails.  The quark-hadron phase transition takes place when  equations (\ref{e41}) and (\ref{e4-1}) become equal (the Gibbs criteria)
\begin{align}\label{e4-3}
\alpha T_{c}^{8}+\beta T_{c}^{4}+\gamma_{T}T_{c}^{2}+B=0,
\end{align}
where $\alpha=\frac{\pi^{2}}{30T_{int}^{4}}$ and $\beta=(\frac{\pi^{2}}{30}-a_{q}-\alpha_{T})$. If we take $m_{s}=200 MeV$ and $B^{1/4} = 200 MeV$,  equation (\ref{e4-3}) will have two negative, four pure imaginary and two real positive solutions of the order $T_{c}\approx120MeV$ and $T_{c}\approx340MeV$, which according to the predictions of QCD concerning deconfinement-confinement phase transition in the early universe, only $T_{c}\approx120MeV$ is admissible. It is worth mentioning at this stage  that the order of phase transition is given by the emergence of singularities (non-analyticities) in functions representing physical quantities; if the energy density $\rho$ or entropy density $s$ of a system in passing through the critical temperature $T_{c}$ displays a discontinuity then the transition is of first order, otherwise the transition is said to be of second order or crossover. Hence the phase transition given by  equations (\ref{e41}), (\ref{e40}), (\ref{e4-1}) and (\ref{e4-2}) at $T_{c}=120MeV$ is first order with energy and entropy density discontinuities respectively
\begin{align}
\Delta\rho(T_{c})\equiv\rho_{q}(T_{c})-\rho_{h}(T_{c})\approx 1.25B,\quad \Delta s(T_{c})\equiv s_{q}(T_{c})-s_{h}(T_{c})\approx0.1B.
\end{align}

\section{Non-causal fluid and dynamical evolution of HL universe during first order transition}
Observations show that the universe is not filled with an ideal  perfect fluid and thus the viscosity plays an important role in the evolution of the universe \cite{50}. In what follows we assume that the quark-hadron phase transition mechanism takes place in a viscous medium  preserving isotropic and homogeneous conditions. Indeed the early universe is assumed to have been filled with a bulk viscous cosmological fluid  which can be described by a FRW metric (\ref{em5}). Note that during the quark-hadron phase transition, the cosmic background fluid was composed of a hot viscous soap of particles such as leptons and photons, undergoing electro-weak transition. Hence viscosity effects  play an important role in the EoS of the quark-hadron phase together with background geometry of the early cosmological settings. A point of caution is that if the quark-hadron phase transition  takes place in a viscous medium with anisotropic space then  in addition to bulk viscosity,  shear viscosity $\eta$ should also be taken into account.

At temperatures greater than  $10^{16}GeV$, the thermal history of evolution of the early universe suggests that the universe cannot be regarded as an equilibrium system \cite{N.C}.  The bulk viscosity is regarded as the value of the pressure needed to restore equilibrium to the  system being expanded\cite{S}. The grand unified theory (GUT) predicts that the existence of bulk viscosity may lead to an inflationary cosmology. In  \cite{59}  an inflationary Bianchi type model in the presence of both bulk and shear viscosity is studied. Also in  \cite{60}, by taking into account the bulk viscosity in Brans-Dicke theory, the authors discover an accelerated phase for the universe. In \cite{S},  the effects of bulk viscosity on the early evolution of the universe is studied within the frame work of standard  cosmology with an EoS of the form  $p=(\gamma-1)\rho$,  leading to an inflationary universe with radiation-dominated phases.  Before going any further, one should introduce an evolution equation for the bulk viscosity. We assume that the bulk viscosity coefficient depends on $\rho$ via a power-law of the form
\begin{align}\label{e43}
\xi(\rho)=\xi_{0}\rho^{n}.
\end{align}
where $\xi_{0}$ is a non-negative dimensional constant and  $n$ is a non-negative numerical constant (in follow set $\xi_{0}=1$). For the rest of the discussion we use natural units with $ c=\hbar=k_{B}=1$ where $8\pi G=t_{Pl}^{2}=1$, so that the unit of the reduced Planck time $t_{Pl}$ is $MeV^{-1}$.

\begin{figure}
\includegraphics[width=8 cm]{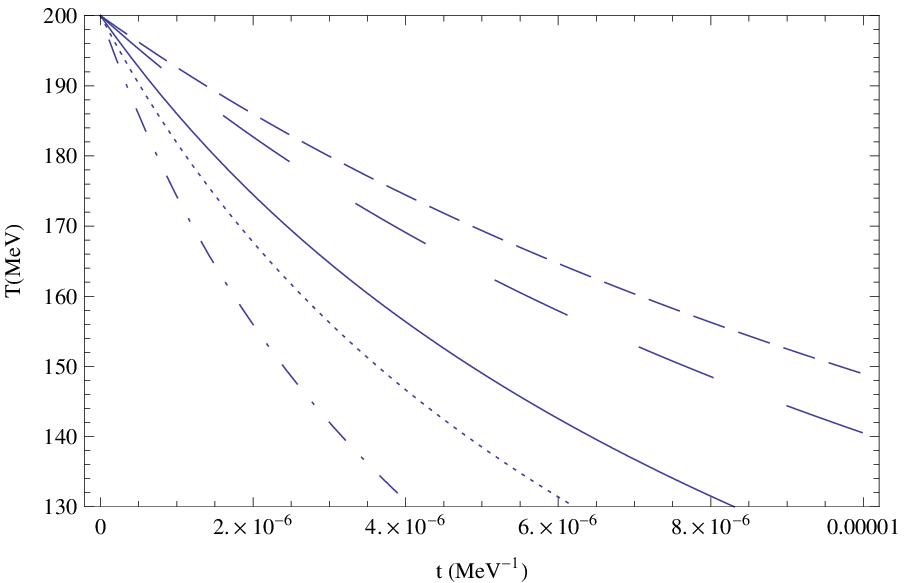}
\centering\caption{Numerical solution of temperature $T$ during the quark phase as a function
of $t$ for $n=0.2$, $B^{1/4}=200MeV$ and different values of $\lambda$; $\lambda=3$ (short dashed curve), $\lambda=2$ (long dashed curve), $\lambda=1$ (solid curve), $\lambda=0.7$ (dotted curve), $\lambda=0.5$ (dotted-dashed curve), respectively. The
background fluid of HL universe is characterized by non-causal Eckart theory.{\label{fig.1} \small}}
\end{figure}

For $T > T_{c}$ before the phase transition, the universe is in  pure quark phase. To first order deviation from a perfect fluid we take the Eckart fluid (\ref{e3}) together with relations (\ref{e17}),  (\ref{e43}) and the Friedmann equation (\ref{e15}) ($k=\Lambda=0$). Substitution of the equations of state of the quark matter (\ref{e41}) using the self-interacting potential (\ref{e40}), finally  leads to
\begin{widetext}
\begin{align}\label{e44}
\frac{dT}{dt}=-\frac{a_{2}T^{4}\sqrt{a_{1}T^{4}+\gamma_{T}T^{2}+B}}{(3a_{1}T^{3}+2\gamma_{T}T)}+
\frac{a_{3}(a_{1}T^{4}+\gamma_{T}T^{2}+B)^{n+1}}{(3a_{1}T^{3}+2\gamma_{T}T)},
\end{align}
\end{widetext}
where
\begin{align}
a_{1}=3a_{q}-\alpha_{T},\quad a_{2}=\frac{12a_{q}}{\sqrt{2(3\lambda-1)}} , \quad a_{3}=\frac{3}{(3\lambda-1)}.
\end{align}
Equation (\ref{e44}) does not have an exact  solution and  may be solved numerically, with the result  presented in figure (\ref{fig.1}), showing  the behavior of temperature as a function of cosmic time $t$ for different values of $\lambda$ with $n=0.2$. The value of $n$ is bounded to the interval $0\leq n<0.5$.
As mentioned earlier, the values of $\lambda$ is restricted close to $\lambda=1$. However, in order to show  the explicit dependence of cosmological parameters to the running coupling constant of HL gravity $\lambda$, we use  an open window around $\lambda=1$.
As one can see from figure (\ref{fig.1}), for $1/3<\lambda<1$ the temperature is much lower than the standard cosmology ($\lambda=1$) while for $\lambda>1$ it is much higher. One  will notice that \cite{F.M.H} the overall behavior of temperature in the absence of bulk viscosity is the same as that in figure \ref{fig.1}.

\begin{figure}
\includegraphics[width=8 cm]{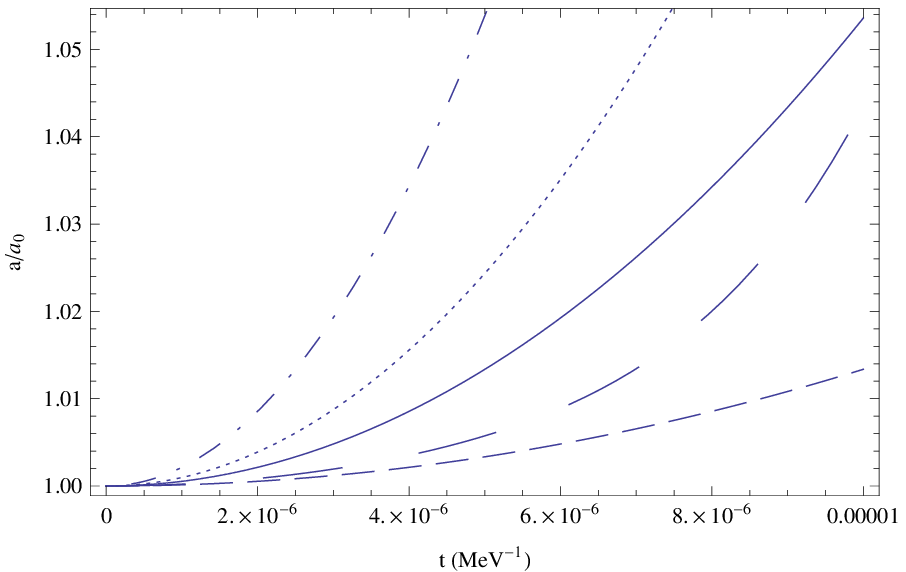}
\centering\caption{Numerical solutions of the dimensionless scale factor $a/a_{0}$ during the quark phase as a function
of $t$ for $n=0.4$, $B^{1/4}=200MeV$ and different values of $\lambda$; $\lambda=3$ (short dashed curve), $\lambda=2$ (long dashed curve), $\lambda=1$ (solid curve), $\lambda=0.7$ (dotted curve), $\lambda=0.5$ (dotted-dashed curve), respectively. The
background fluid of HL universe is characterized by non-causal Eckart theory.{\label{fig.a1} \small}}
\end{figure}

Another cosmological parameter of the HL early Universe that we are interested in is the scale factor $a$. Substituting (\ref{e43}), $\Pi=-3\xi H$ in the second Friedmann equation (\ref{e16}) and using the first Friedmann equations (\ref{e15}) and relation $p=\frac{\rho-4B}{3}$, in pure quark phase one  has
\begin{align}\label{ea1}
\frac{d^{2}a}{dt^{2}}+\frac{1}{3a}\left(\frac{da}{dt}\right)^{2}-\frac{\gamma^{n-1}}{2}\frac{1}{a^{2n}}\left(\frac{da}{dt}\right)^{2n+1}-\eta a=0,
\end{align}
where $\gamma=\frac{(3\lambda-1)^{n-1}}{2^{n}}$ and $\eta=\frac{4B}{3(3\lambda-1)}$. Numerical solutions of (\ref{ea1}) for different values of $\lambda$ and $n=0.4$ are shown in figure \ref{fig.a1}.
Clearly, observed that in viscous QGP phase  filled with Eckart fluid, for $1/3<\lambda<1$, the evolution of the scale factor $a$ is faster than the standard cosmology ($\lambda=1$) while for $\lambda>1$ it is slower.

\begin{figure}
\includegraphics[width=8 cm]{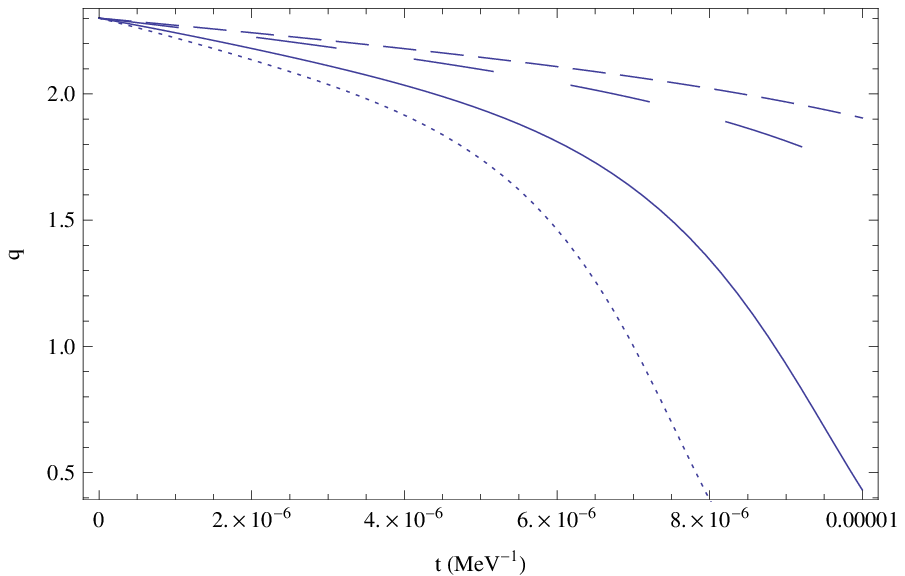}
\centering\caption{Numerical solution of deceleration parameter $q$ during the quark phase as a function
of $t$ for $n=0.2$, $B^{1/4}=200MeV$ and different values of $\lambda$; $\lambda=3$ (short dashed curve), $\lambda=2$ (long dashed curve), $\lambda=1$ (solid curve), $\lambda=0.7$ (dotted curve), respectively. The
background fluid of HL universe is characterized by non-causal Eckart theory.{\label{fig.2} \small}}
\end{figure}

The deceleration parameter can be calculated using
\begin{align}\label{e45}
q=\frac{dH^{-1}}{dt}-1=-\frac{\dot{H}}{H^{2}}-1,
\end{align}
which, upon using  Friedmann equations (\ref{e15}), (\ref{e16}) gives
\begin{align}\label{e46}
q=\frac{3}{2}\left(1+\frac{p}{\rho}\right)-\sqrt{\frac{a_{3}\rho^{2n-1}}{2}}.
\end{align}
It is worth mentioning that one may classify uniform universes according to the values of $H$ and $q$. Such a classification should be called ``kinematic classification,'' in contrast to a classification in terms of the curvature, which is a geometric classification. Kinematically, uniform universes fall into several classes of which  only three sets are possible candidates for our universe at present;  $q>0$, $H>0$  corresponding to standard decelerating models whereas  $q<0$ indicates accelerated expansion and $q=0$ describe a universe expanding with zero deceleration. Using the EoS of the quark phase, equation (\ref{e41}) together with (\ref{e40}), one obtains
\begin{align}\label{e47}
q=\frac{3(a_{1}+b_{1})T(t)^{4}}{2(a_{1}T(t)^{4}+\gamma_{T}T(t)^{2}+B)}-\sqrt{\frac{a_{3}(a_{1}T(t)^{4}+\gamma_{T}T(t)^{2}+B)^{2n-1}}{2}},
\end{align}
where $b_{1}=a_{q}+\alpha_{T}$. Using equation (\ref{e44}) one may observe the behavior $q$ in term of cosmic tim $t$, see figure \ref{fig.2}.
Figure \ref{fig.2} shows that the deceleration parameter is positive and decrease with the cosmic time $t$, that is, an HL universe in  viscous pure quark phase is undergoing decelerated expansion. The range of $n$ for which the deceleration parameter is positive is $0\leq n<0.5$. Also it shows that for $1/3<\lambda<1$ it drops faster than that in  standard cosmology ($\lambda=1$) while for $\lambda>1$ its drop is slower than the standard cosmology.

\begin{figure}
\includegraphics[width=8 cm]{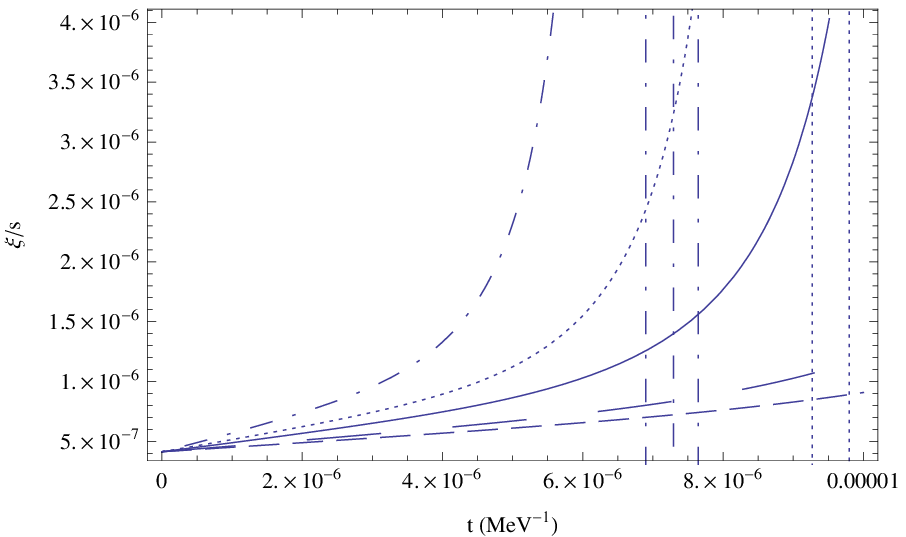}
\centering\caption{Numerical solution of dimensionless ratio $\xi/s$ during the quark phase as a function
of $t$ for $n=0.2$, $B^{1/4}=200MeV$ and different values of $\lambda$; $\lambda=3$ (short dashed curve), $\lambda=2$ (long dashed curve), $\lambda=1$ (solid curve), $\lambda=0.7$ (dotted curve), $\lambda=0.5$ (dotted-dashed curve), respectively. The
background fluid of HL universe is characterized by non-causal Eckart theory.{\label{fig.3} \small}}
\end{figure}

The last thermodynamical quantity  of interest is the dimensionless ratio of the bulk viscosity coefficient to entropy density, $\frac{\xi}{s}$. Using relations \,(\ref{e41}) , \,(\ref{e40}), \,(\ref{e43}) and the definition of entropy density, $s(T)=\frac{dp}{dT}$, we obtain
\begin{align}\label{e48}
\frac{\xi}{s}=\frac{(a_{1}T(t)^{4}+\gamma_{T}T(t)^{2}+B)^{n}}{(4b_{1}T(t)^{3}-2\gamma_{T}T(t))}.
\end{align}
The behavior of this ratio in terms of $t$ is shown in figure \ref{fig.3}. It clearly shows the growth  of $\frac{\xi}{s}$ with cosmic time after the  start of QGP-hadron phase transition corresponding to the downfall of effective temperature. Near the critical temperature $T_{c}$ its value grows rapidly.  In general, the behavior in figure (\ref{fig.3}) is in agreement with lattice QCD results as has been reported in \cite{H.B}. Also, a glance at figure 1 in \cite{J.R} and  figure 3 in \cite{J.R.V} reveals that at the QGP phase away from $T_{c}$, the value of $\xi/s$  falls from a maximum value toward zero, in agreement with our model.

\begin{figure}
\includegraphics[width=8 cm]{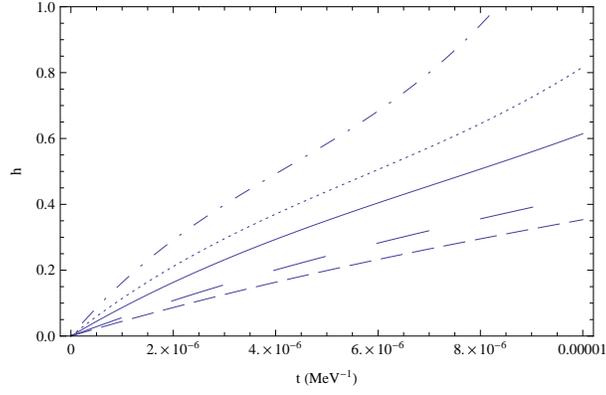}
\centering\caption{Numerical solution of hadron fraction $h$ during the quark-hadron phase transition as a function
of $t$  for $n=0.2$ and different values of $\lambda$; $\lambda=3$ (short dashed curve), $\lambda=2$ (long dashed curve), $\lambda=1$ (solid curve),
$\lambda=0.7$ (dotted curve), $\lambda=0.5$ (dotted-dashed curve), respectively. The
background fluid of HL universe is characterized by non-causal Eckart theory.{\label{fig.4} \small}}
\end{figure}

\begin{figure}
\includegraphics[width=8 cm]{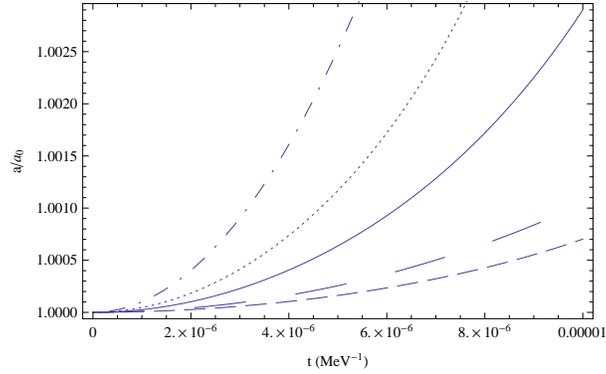}
\centering\caption{Numerical solution of dimensionless scale factor $a/a_{0}$ during the quark-hadron phase transition as a function
of $t$ for $n=0.4$ and different values of $\lambda$; $\lambda=3$ (short dashed curve), $\lambda=2$ (long dashed curve), $\lambda=1$ (solid curve),
$\lambda=0.7$ (dotted curve), $\lambda=0.5$ (dotted-dashed curve), respectively. The
background fluid of HL universe is characterized by non-causal Eckart theory.{\label{fig.a2} \small}}
\end{figure}

For $T = T_{c}$, during the phase transition, temperature and pressure are constant and thereupon the entropy density $s$
is conserved. Also, during the phase transition $\rho(t)$ decreases from $\rho_{q}(T_{c})= \rho_{Q}$ to $\rho_{h}(T_{c})= \rho_{H}$. For phase transition temperature of $T_{c} =120 MeV$ we have $\rho_{Q}\approx 4.5\times 10^{9}MeV^{4}$ and $\rho_{H}\approx 4\times 10^{8}MeV^{4}$, respectively. Also the value of the pressure of the cosmological fluid during the phase transition is $p_{c}\approx 10^{8}MeV^{4}$. Following the first reference in \cite{55}, we replace $\rho(t )$ by $h(t)$, so that the volume fraction of matter in the hadron phase is given by
\begin{align}\label{e49}
\rho(t)=\rho_{h}(t)+\rho_{q}(1-h(t))=\rho_{q}(1+mh(t)),
\end{align}
where $m=\frac{\rho_{H}}{\rho_{Q}}-1$. The beginning of the phase transition is characterized by $h(t_{c})= 0$ where $t_{c}$ is the time for which $t_{c} \equiv\rho_{Q}$, while the end of the transition is characterized by $h(t_{h})=1$ with $t_{h}$ being the time signaling the end and corresponding to $\rho(t_{h}\equiv\rho_{h})$. For $t>t_{h}$ the universe enters into the hadronic phase. Now, substituting equation (\ref{e49}) in the energy-momentum  conservation equation leads to
\begin{align}\label{e50}
\frac{dh}{dt}=-\sqrt{a_{3}\rho_{Q}(1+mh)}(r+h)+\frac{3a_{3}(\rho_{Q}(1+mh))^{n+1}}{(\rho_{H}-\rho_{Q})},
\end{align}
where $r=\frac{\rho_{Q}+p_{c}}{\rho_{H}-\rho_{Q}}$. Figure \ref{fig.4} shows  variation of the hadron fraction $h(t)$
as a function of cosmic time $t$ for different values of the running coupling constant $\lambda$ with $n = 0.2$.
Again, the numerical values of $n$ are restricted to the interval $0\leq n<0.5$. Figure \ref{fig.4} implies that in analogy to \cite{F.M.H}, the hadron fraction in the context of HL gravity for $1/3<\lambda<1$ is much higher than the standard cosmology ($\lambda=1$) while for $\lambda>1$ it is much lower. In a similar fashion, the scale factor $a$ appearing in HL cosmology during viscous quark-hadron phase transition obeys the differential equation
\begin{align}\label{ea2}
\frac{d^{2}a}{dt^{2}}+\frac{1}{6a}(\frac{da}{dt})^{2}-\frac{\gamma^{n-1}}{2}.\frac{1}{a^{2n}}(\frac{da}{dt})^{2n+1}-\frac{p_{c}}{2\gamma} a=0.
\end{align}
The numerical solutions of (\ref{ea2}) for different values of $\lambda$ are shown in figure \ref{fig.a2}.

As can be seen, during viscous quark-hadron phase transition, the evolution of scale factor $a$ is slower than viscous QGP phase. As in the viscous QGP phase, for $1/3<\lambda<1$, the scale factor has higher values than the standard cosmology ($\lambda=1$) while for $\lambda>1$ it has lower values. In a mixed quark-hadron phase, since the temperature and pressure are constant, one has  $\xi/s \rightarrow \infty$.

\begin{figure}
\includegraphics[width=8 cm]{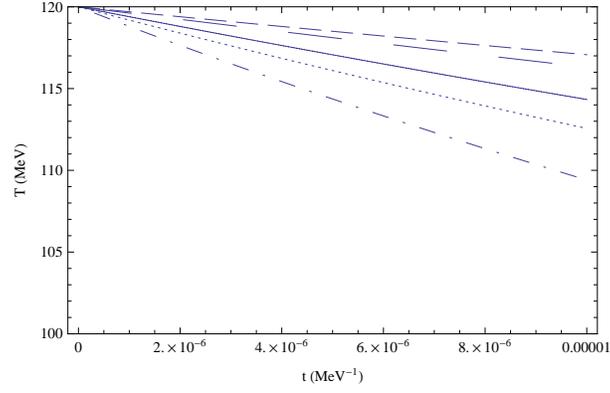}
\centering\caption{Numerical solution of temperature $T$ during the hadronic phase as a function
of $t$  for $n=0.2$ and different values of $\lambda$: $\lambda=3$ (short dashed curve), $\lambda=2$ (long dashed curve), $\lambda=1$ (solid curve),
$\lambda=0.7$ (dotted curve), $\lambda=0.5$ (dotted-dashed curve), respectively. The
background fluid of HL universe is characterized by non-causal Eckart theory.{\label{fig.5} \small}}
\end{figure}

For $T < T_{c}$,  using relations (\ref{e15}), (\ref{e17}) and (\ref{e3}) which represent a first order deviation from perfect fluid  and substituting (\ref{e43}) and the EoS of the hadron matter obtained from the low temperature expansion approach, i.e.  equations (\ref{e4-1}) and (\ref{e4-2}) we find
\begin{widetext}
\begin{align}\label{e50}
\frac{dT}{dt}=-\sqrt{\frac{a_{3}(d_{1}T(t)^{4}+d_{2}T(t)^{8})}{6}}\left(\frac{28d_{1}T(t)^{4}+
24d_{2}T(t)^{8}}{42d_{1}T(t)^{3}+168d_{2}T(t)^{7}}\right)+
\frac{9a_{3}(d_{1}T(t)^{4}/9+3d_{2}T(t)^{8}/7)^{n+1}}{(2d_{1}T(t)^{3}+8d_{2}T(t)^{7})},
\end{align}
\end{widetext}
where $d_{1}=9\pi^{2}/10$ and $d_{2}=63\pi^{2}/30T_{int}^{4}$. Variation of temperature in terms of the cosmic time for the hadronic phase for different values of running coupling constant $\lambda$ with $n=0.2$ is presented in figure \ref{fig.5}.
Figure \ref{fig.5} implies that, in the viscous hadronic phase, the running coupling constant  dependence on temperature is similar to the viscous QGP phase.
As before,  the behavior of the scale factor in the pure hadronic phase should be determined by an explicit relationship between $p$ and $\rho$. However,  unlike the QGP phase,  the EoS equations (\ref{e4-1}) and (\ref{e4-2}) are not well defined. Besides, in the viscous hadronic phase  away from the critical temperature $T_{c}\approx120MeV$ and the ratio $(T/T_{int})^{4}$ should be negligible. Therefore, it is reasonable to use the approximation $p\sim\frac{\rho}{3}$ which, in a same manner as before, leads to
\begin{align}\label{ea3}
\frac{d^{2}a}{dt^{2}}+\frac{1}{3a}\left(\frac{da}{dt}\right)^{2}-\frac{\gamma^{n-1}}{2}\frac{1}{a^{2n}}\left(\frac{da}{dt}\right)^{2n+1}=0.
\end{align}
\begin{figure}
\includegraphics[width=8 cm]{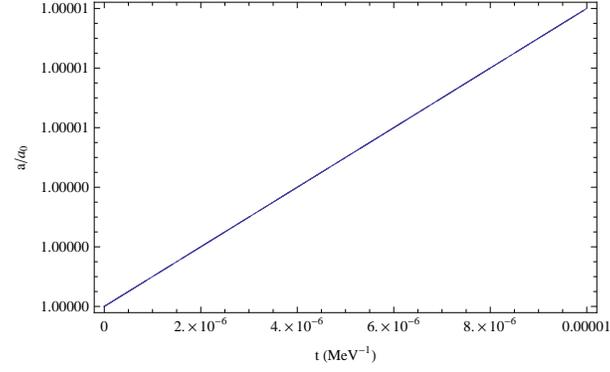}
\centering\caption{Numerical solution of dimensionless scale factor $a/a_{0}$ during the hadronic phase as a function
of $t$ for $n=0.4$ and different values of $\lambda$; $\lambda=3$ (short dashed curve), $\lambda=2$ (long dashed curve), $\lambda=1$ (solid curve),
$\lambda=0.7$ (dotted curve), $\lambda=0.5$ (dotted-dashed curve), respectively. The
background fluid of HL universe is characterized by non-causal Eckart theory.{\label{fig.a3} \small}}
\end{figure}
As one can see from figure \ref{fig.a3}, for the viscous hadronic phase, the evolution of the scale factor is too slow and indistinguishable. It is interesting to note that in crossing the QGP phase and approaching the hadronic phase, the growth of the scale factor becomes slower.
To obtain the deceleration parameter we substitute the EoS of the hadron matter (\ref{e4-1}) in (\ref{e46}) and find
\begin{align}\label{e51}
q=\frac{14d_{1}T(t)^{4}+12d_{2}T(t)^{8}}{7(d_{1}T(t)^{4}+d_{2}T(t)^{8})}-\sqrt{\frac{a_{3}(d_{1}T(t)^{4}/9+
d_{2}T(t)^{8}/9)^{2n-1}}{2}}.
\end{align}
Figure \ref{fig.6} shows the general behavior of $q$ given by equation (\ref{e51}) as a function of the cosmic time. One observes that in the viscous hadronic phase,  variation of the deceleration parameter $q$ is too slow.  For the interval $0\leq n\leq 0.5$, $q$ is positive and for  $0.5<n\leq1$ it is negative and unlike the viscous QGP phase, for $q>0$, the variation is very slow.

\begin{figure}
\includegraphics[width=8 cm]{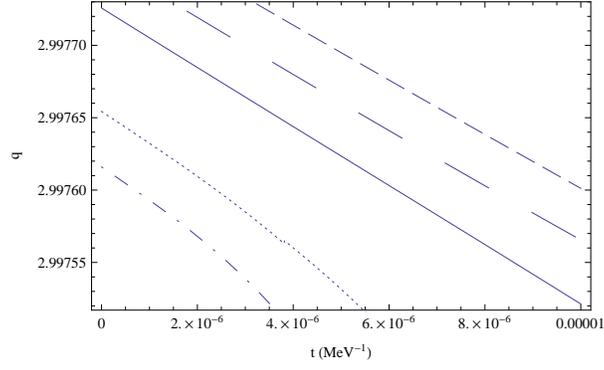}
\centering\caption{Numerical solution of deceleration parameter $q$ during the hadronic phase as a function
of $t$ for $n=0.2$ and different values of $\lambda$; $\lambda=3$ (short dashed curve), $\lambda=2$ (long dashed curve), $\lambda=1$ (solid curve),
$\lambda=0.7$ (dotted curve), $\lambda=0.5$ (dotted-dashed curve), respectively. The
background fluid of HL universe is characterized by non-causal Eckart theory.{\label{fig.6} \small}}
\end{figure}
Finally, using the EoS for hadron matter (\ref{e4-1}) in $\xi/s$ we obtain
\begin{align}\label{e52}
\frac{\xi}{s}=\frac{189(d_{1}T(t)^{4}/9+d_{2}T(t)^{8}/9)^{n}}{(28d_{1}T(t)^{3}+48d_{2}T(t)^{7})}.
\end{align}
\begin{figure}
\includegraphics[width=8 cm]{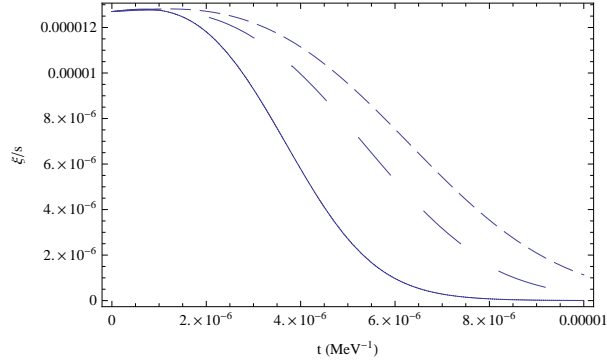}
\centering\caption{Numerical solution of dimensionless ratio $\xi/s$ during the hadronic phase as a function
of $t$  for $n=0.2$ and different values of $\lambda$; $\lambda=3$ (short dashed curve), $\lambda=2$ (long dashed curve), $\lambda=1$ (solid curve),
respectively. The background fluid of HL universe is characterized by non-causal Eckart theory.{\label{fig.7} \small}}
\end{figure}

Figure \ref{fig.7} shows the behavior of this ratio with respect to the cosmic time. As the time grows and temperature decreases, the ratio $\xi/s$  falls off a maximum value toward zero for  $0\leq n\leq0.8$. Here too, there exists a good qualitative agreement with lattice QCD results. In \cite{J.W} the authors calculate the bulk viscosity $\xi$ using  Boltzmann equation with  kinetic theory generalized to incorporate the trace anomaly and show a monotonic increase for dimensionless ratio $\xi/s$ below $T = 120 MeV$, the reader is referred to  figure 1 in \cite{J.W}. Also in the second reference in \cite{J.R.V}, figure 8  the author, using the resonance gas model including Hagedorn states \cite{J.N.H}, obtains the same behavior for  $\xi/s$ in the interval $100 MeV\leq T<200 MeV$.  As the final word in this section we note that the evolution of  physical quantities such as temperature $T$, deceleration parameter $q$ and dimensionless ratio $\frac{\xi}{s}$ is best described by the value of $n$ in a power-law bulk viscosity given by $\rho^{n}$ where $0\leq n<0.5$.

\section{Causal fluid and dynamical evolution of HL universe during first order phase transition}
\begin{figure}
\includegraphics[width=8 cm]{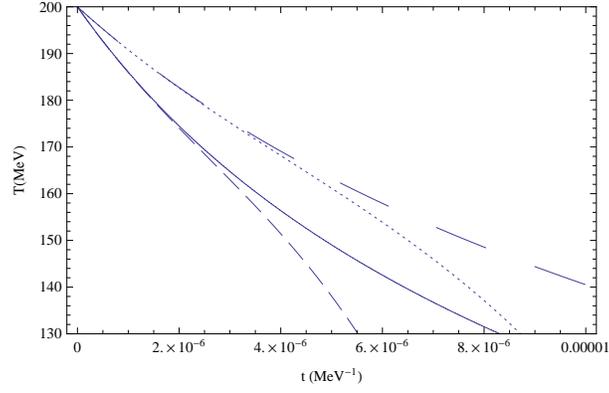}
\centering\caption{Numerical solution of temperature $T$ during the quark phase as a function
of $t$. The solid ($\lambda=1$) and long dashed curves ($\lambda=2$) shows the results when the background geometry is characterized by Eckart fluid. The short dashed ($\lambda=1$) and dotted curves ($\lambda=2$) shows the results when the background geometry is assumed to
be filled with an steady full version of the IS fluid. {\label{fig.8} \small}}
\end{figure}

The approximate nature of the relativistic Eckart fluid is of first order in the sense that it only considers first-order deviations from equilibrium and hence  could lead to non-causal and unstable problems. To remedy such problems additional terms should be used to convert the equations governing dissipative quantities from  algebraic first-order type into differential evolution equations. Following Eckart's work  a relativistic second-order theory was presented by Israel and  Stewart which we refer to as the IS theory.  It has two versions, truncated and full causal equations. Results show that the truncated causal thermodynamics of the bulk viscosity like non-causal Eckart theory leads to unobserved behavior in the late universe, while  solutions of the full causal theory are well behaved for all times \cite{2}. Hence in this section, to study the phenomena of first order phase transition in non-detailed balance early HL, we will consider the full evolution equations of the IS theory (\ref{e9}). Here we assume that the isotropic and homogeneous background cosmological fluid is an IS one which, in effect, represents  seconded order deviations from equilibrium. For the case of steady evolution,  equations (\ref{e9}) reduce to
\begin{align}\label{e52}
\Pi=-3\xi_{eff} H,\quad \xi_{eff}=\frac{\xi}{1+3/2\tau H}.
\end{align}
It is  worth recalling that the phenomenological model described by$\tau=\xi/\rho$ has been adopted as one way of ensuring that viscous signals do not exceed the speed of light in the truncated theory\cite{D.P.B, 2}.
Substituting the above relation and bulk viscosity coefficient (\ref{e43}) in the continuity equation (\ref{e17}) and  using the first Friedmann equation (\ref{e15}) for $k=\Lambda=0$ leads to
\begin{align}\label{e53}
\dot{\rho}=-A(\rho^{3/2}+p\rho^{1/2})+\frac{27A^{2}\rho^{n}}{27+13.5A\rho^{n-1/2}},
\end{align}
where $A=\frac{3}{\sqrt{3(3\lambda-1)}}$.
As one can see for $T > T_{c}$,  substituting the EoS equation of the quark matter (\ref{e41}) together with self-interaction potential term (\ref{e40}) in (\ref{e53}) results in a relation giving the variation of temperature with respect to cosmic time in the presence of the causal cosmological background fluid
\begin{widetext}
\begin{align}\label{e54}
\frac{dT}{dt}=-\frac{A(a_{1}T^{4}+\gamma_{T}T^{2}+B)^{3/2}+(b_{1}T^{4}-\gamma_{T}T^{2}-B)
(a_{1}T^{4}+\gamma_{T}T^{2}+B)^{1/2}}{(4a_{1}T^{3}+2\gamma_{T}T)}\nonumber\\ +\frac{27A^{2}(a_{1}T^{4}+\gamma_{T}T^{2}+B)^{n}}{27(4a_{1}T^{3}+2\gamma_{T}T)+
13.5A(a_{1}T^{4}+\gamma_{T}T^{2}+B)^{n-1/2}(4a_{1}T^{3}+2\gamma_{T}T)},
\end{align}
\end{widetext}
which may be solved numerically for different values of $\lambda$. The different behaviors  of the steady truncated (Eckart) and full version of the IS fluid in viscous QGP phase are shown in figure \ref{fig.8}.
We see that  variation of temperature $T$ in the presence of IS fluid is independent of the values of $n$. As one can see from figure \ref{fig.8}, curves belong to the IS fluid are lower than Eckart fluid. In other words, in viscous QGP phase the temperature of HL early universe filled with full version of the IS fluid declines faster than the Eckart fluid.

\begin{figure}
\includegraphics[width=8 cm]{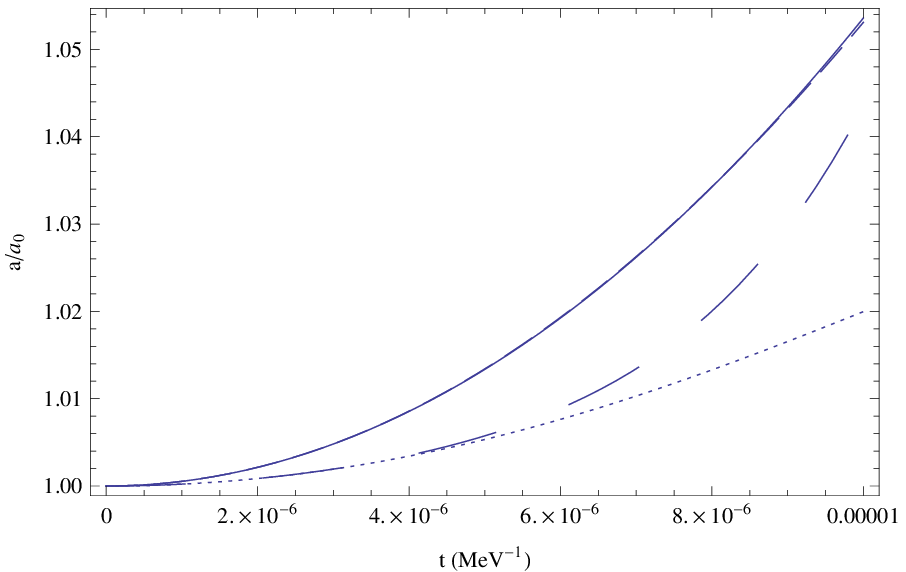}
\centering\caption{Numerical solution of dimensionless scale factor $a/a_{0}$ during the quark phase as a function
of $t$ for $B^{1/4}=200MeV$. The solid ($\lambda=1$) and long dashed curves ($\lambda=2$) show the results when the background geometry is characterized by Eckart fluid. The short dashed ($\lambda=1$) and dotted curves ($\lambda=2$) show the results when the background geometry is assumed to
be filled with an steady full version of the IS fluid. The curves of fluids for $\lambda=1$ are indistinguishable.{\label{fig.a4} \small}}
\end{figure}

 Now, using (\ref{e52}) and  Friedmann equations (\ref{e15}) and (\ref{e16}), in the presence of steady full version of the IS fluid, we have
\begin{align}\label{ea4}
\frac{d^{2}a}{dt^{2}}+\frac{1}{3a}\left(\frac{da}{dt}\right)^{2}-\frac{12\left(\frac{da}{dt}\right)^{2n+1}}{(2\gamma)^{1-n}a^{2n}+3\times2^{2-n}\left(\frac{da}{dt}\right)^{2n-1}}-\eta a=0.
\end{align}
 Although the general behavior  of the above  equation is similar to \ref{fig.a1},  to comparer it to Eckart and IS fluid, we have shown it in figure \ref{fig.a4}. One can see that for standard cosmology ($\lambda=1$) curves representing the Eckart and steady full version of the IS fluid are indistinguishable, but for $\lambda\neq1$, the  Eckart curve is higher than that of the IS fluid.

\begin{figure}
\includegraphics[width=8 cm]{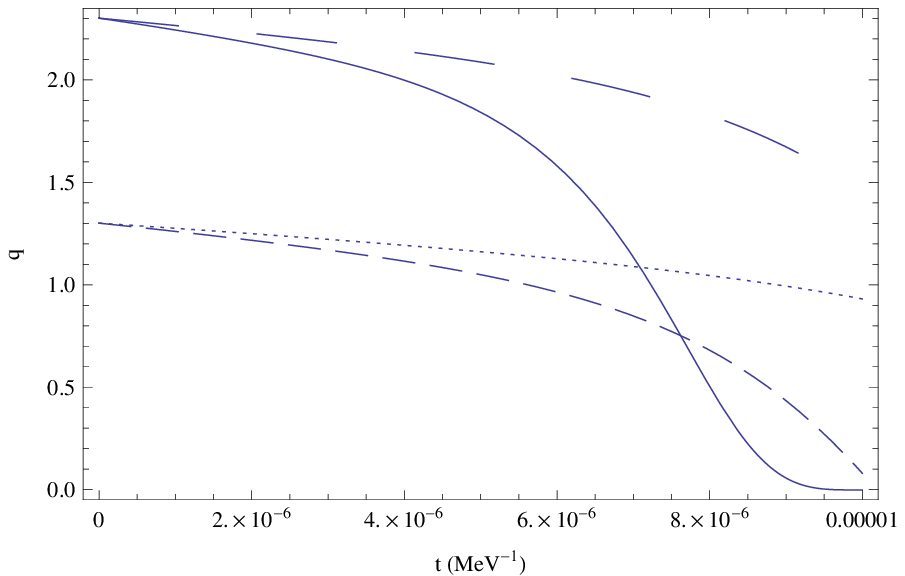}
\centering\caption{Numerical solution of deceleration parameter $q$  as a function
of $t$ for $n=0.2$ and $B^{1/4}=200MeV$. The solid ($\lambda=1$) and long dashed curves ($\lambda=2$) show the results when the background geometry is characterized by Eckart fluid. The short dashed ($\lambda=1$) and dotted curves ($\lambda=2$) show the results when the background geometry is assumed to
be filled with an steady full version of the IS fluid. {\label{fig.9} \small}}
\end{figure}

Let us now calculate the deceleration parameter $q$ using relations (\ref{e43}), (\ref{e45}), (\ref{e52}) and the Friedmann equation (\ref{e15}), finding
\begin{align}\label{e55}
q=\frac{3}{2}\left(1+\frac{p}{\rho}\right)-\frac{3\rho^{n}}{\sqrt{2/3}A\rho+\rho^{n}}.
\end{align}
Now, using the EoS we have
\begin{align}\label{e56}
q=\frac{3(a_{1}+b_{1})T(t)^{4}}{2(a_{1}T(t)^{4}+\gamma_{T}T(t)^{2}+B)}-
\frac{3(a_{1}T(t)^{4}+\gamma_{T}T(t)^{2}+B)^{n}}{\sqrt{2/3}A(a_{1}T(t)^{4}+
\gamma_{T}T(t)^{2}+B)+(a_{1}T(t)^{4}+\gamma_{T}T(t)^{2}+B)^{n}}.
\end{align}
We see that $q$ has two types of behavior depending on whether $0\leq n<0.5$, $q>0$, or $n\geq0.5$,  $q<0$. In figure \ref{fig.9} we have compared the behavior of $q$ in HL early universe filled with Eckart fluid with that of the steady full version of the IS fluid. It is seen that curves belonging to the steady full version of the IS fluid are lower than the Eckart fluid.
In a fashion analogous to the previous section, we consider the dimensionless ratio $\frac{\xi}{s}$. Using the EoS (\ref{e39}) and (\ref{e40}) we solve equation (\ref{e52}) numerically with results too similar to figure \ref{fig.3}.  Here we also see a good qualitative agreement with lattice QCD calculation for $0\leq n\leq0.8$.
For $T = T_{c}$, using equations (\ref{e53}) and (\ref{e40}) we find
\begin{align}\label{e57}
\frac{dh}{dt}=\frac{-A((\rho_{Q}(1+mh(t)))^{3/2}+p_{c}(\rho_{q}(1+mh(t)))^{1/2})}{m\rho_{Q}}+
\frac{27A^{2}(\rho_{Q}(1+mh(t)))^{n}}{(27m\rho_{Q}+
13.5Am\rho_{Q}(\rho_{Q}(1+mh(t)))^{n-1/2})}.
\end{align}
\begin{figure}
\includegraphics[width=8 cm]{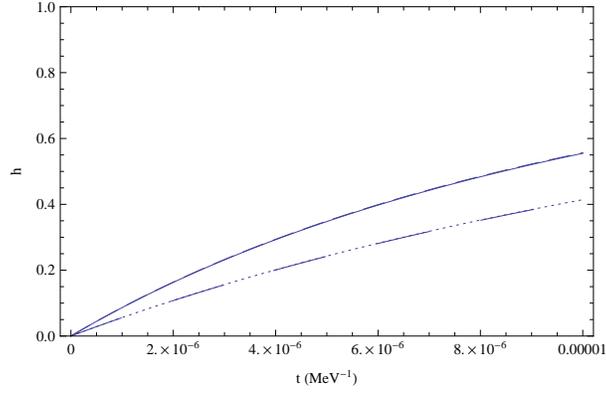}
\centering\caption{Numerical solution of hadron fraction $h$ during the quark-hadron phase transition as a function
of $t$. The solid ($\lambda=1$) and long dashed curves ($\lambda=2$) shows the results when the background geometry is characterized by Eckart fluid. The short dashed ($\lambda=1$) and dotted curves ($\lambda=2$) shows the results when the background geometry is assumed to
be filled with an steady full version of the IS fluid. The curves of both fluids for different values of $\lambda$ are indistinguishable.{\label{fig.10} \small}}
\end{figure}
In figure \ref{fig.10} we have investigated the behavior of the fraction of hadron $h$ as a function of time for both fluids. In analogy with temperature in viscous  QGP phase, the variation of fraction of hadron is independent of numerical values of $n$. As can be seen, for different values of $\lambda$, the behavior of curves in both fluids, are matched and indistinguishable.
During phase transition, scale factor $a$ of HL early Universe filled with steady full version of the IS fluid, will obey the differential equation
\begin{align}\label{ea5}
\frac{d^{2}a}{dt^{2}}+\frac{1}{6a}\left(\frac{da}{dt}\right)^{2}-\frac{12(\frac{da}{dt})^{2n+1}}{(2\gamma)^{1-n}a^{2n}+
3\times2^{2-n}(\frac{da}{dt})^{2n-1}}+\frac{p_{c}}{3\lambda-1}=0.
\end{align}
We find that its overall behavior is similar to figure \ref{fig.a2}.

\begin{figure}
\includegraphics[width=8 cm]{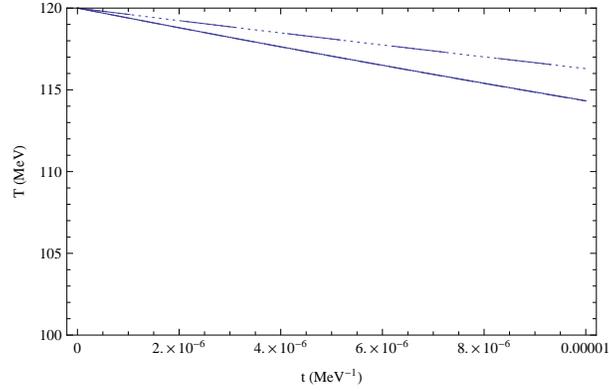}
\centering\caption{Numerical solution of temperature $T$ during the hadronic phase as a function
of $t$ for $B=200^{4}$. The solid ($\lambda=1$) and long dashed curves ($\lambda=2$) show the results when the background geometry is characterized by Eckart fluid. The short dashed ($\lambda=1$) and dotted curves ($\lambda=2$) show the results when the background geometry is assumed to
be filled with an steady full version of the IS fluid. The curves of both fluids for different values of $\lambda$ are indistinguishable.{\label{fig.11} \small}}
\end{figure}

For $T < T_{c}$,  again using the EoS (\ref{e4-2}) and (\ref{e4-3}) in equation (\ref{e53}) we find
\begin{widetext}
\begin{align}\label{e58}
\frac{dT}{dt}&=\frac{1}{2d_{1}T^{3}+8d_{2}T^{7}}\times\nonumber\\
&\left[\frac{27(d_{1}T^{4}+d_{2}T^{8})^{3/2}+9(d_{1}T^{4}+1701d_{2}T^{8})(d_{1}T^{4}+
d_{2}T^{8})^{1/2}}{81}+\frac{243A^{2}(d_{1}T^{4}/9+
d_{2}T^{8}/9)^{n}}{\left(27+13.5A(d_{1}T^{4}/9+d_{2}T^{8}/9)\right)}\right].
\end{align}
\end{widetext}
Figure \ref{fig.11} shows that for  certain values of $\lambda$, the behavior of temperature in both fluids is indistinguishable.  Again, for the hadron phase, the behavior of temperature is independent of numerical values of $n$.

The scale factor $a$ of HL early universe in the presence of steady full version of the IS fluid obeys the differential equation
\begin{align}\label{ea6}
\frac{d^{2}a}{dt^{2}}+\frac{1}{3a}\left(\frac{da}{dt}\right)^{2}-\frac{12(\frac{da}{dt})^{2n+1}}{(2\gamma)^{1-n}a^{2n}+3\times2^{2-n}(\frac{da}{dt})^{2n-1}}=0.
\end{align}
As figure \ref{fig.a3} shows,  here too the growth of the scale factor $a$ is very slow so that curves representing Eckart fluid and steady full version of the IS fluid are indistinguishable.
Now, using the EoS (\ref{e4-1}), (\ref{e4-2}) and substituting the numerical solution of equation (\ref{e53}) in (\ref{e55}) we find a similar behavior for $q$ as a function of  cosmic time $t$, shown in figure \ref{fig.6}.
Finally, in a similar manner, the general behavior of $\xi/s$ in viscous hadronic phase for HL early universe filled with steady full version of the IS fluid is found to be similar to figure \ref{fig.7}. We note that for $0\leq n\leq0.8$ the plot of $\xi/s$ shows a behavior acceptable from the viewpoint of lattice calculations.

It is worth mentioning  that the truncated version of the steady state IS evolution equation reduces to $\Pi=-3\xi H$, i.e. the Eckart fluid. On the other hand it is shown in \cite{2}  that only under condition $|\Pi d/dt(\frac{a^{3}\tau}{\xi T})|\ll \frac{a^{3}H}{T}$ the truncated version is a good approximation to the full causal transport equations. To test this condition, we use the Friedmann equation (\ref{e15})  which  leads to the inequality
\begin{align}\label{e59}
dt\gg\frac{1}{\sqrt{27(3\lambda-1)}}\int\frac{d\rho(T)}{9\rho(T)^{3/2}+\sqrt{3(3\lambda-1)}\rho(T)^{2-n}}+
\frac{1}{\sqrt{72(3\lambda-1)}} \int\frac{dT}{9T\rho(T)^{1/2}+\sqrt{3(3\lambda-1)}T\rho(T)^{1-n}}.
\end{align}
The right hand side is an implicit function of $t$ through $T$. The time interval for which this integral is valid is about  $dt\approx10^{-5}$. Solving the integral (\ref{e59}) numerically for the EoS of the quark-hadron phase for $n=0.2$ and $\lambda=1$ gives a value of the order $10^{-10}$ and $10^{-8}$, respectively. Note that the value of the above integral  depends on the values of $n$, so that increasing  $n$ towards $n<0.5$,  decreases the value and the approximation becomes invalid.

\section{Discussion and summary}
To arrived at a valid understanding of the evolution of the early
universe, we need a reliable EoS for the medium filling out the cosmic background. To this end we adopted the non-detailed balance HL theory in the early universe as a non-equilibrium dissipative system.  Concentrating on the assumption that the bulk viscosity cosmological background fluid obeys the Eckart  evolution equation and the full version of the steady state IS fluid, we studied the dynamical evolution of non-detailed balance HL universe during the  QGP-hadron first order phase transition in the absence of the cosmological constant $\Lambda$ in a flat universe.

In the case of non-causal Eckart fluid we found that for different values of the running coupling constant of HL gravity
$\lambda$, a phase transition occurs. The results show that as in the absence of viscosity \cite{F.M.H}, with a decrease in $\lambda$, the downfall of the effective temperature becomes faster in the viscous QGP and hadronic phase. Also by decreasing $\lambda$, the growth of the scale factor $a$, dimensionless ratio $\xi/s$ and the fall of the deceleration parameter $q$ become faster in the viscous QGP phase. We need to point to the fact that the growth of the scale factor $a$ in the viscous hadronic phase is too slow. Inversely, in the viscous hadron phase $\xi/s$ falls with respect to the cosmic time. It is interesting to point out that in the viscous hadronic phase, this behavior of $\xi/s$ in both phases is qualitatively in agreement with the results of lattice QCD calculations.

As mentioned before, the positive sign of the deceleration parameter $q>0$ in the quark and hadron phase corresponds to decelerating models which  means that the early universe in the QGP and hadron phases underwent a decelerating expansion. Note that in the hadronic phase, the rate of negative acceleration is very slow and almost constant. We also found that the effective temperature has an acceptable behavior for $0<n<0.5$. The non-detailed balance HL world in the QGP phase has a positive deceleration parameter for $0\leq n \leq0.5$. In the hadronic phase however, for $0\leq n \leq0.5$ and  $0.5<n\leq1$ we have $q>0$
and $q<0$ respectively. The dimensionless ratio $\xi/s$ in the QGP and hadron phases for $0\leq n <0.5$ and $0\leq n \leq0.8$ is in line with what one might expect.

Generally speaking, in the presence of the causal full steady state version of the IS relativistic viscous fluid,
variations in terms of $\lambda$ are the same. We showed that the behavior of the effective temperature $T$ in both phases together with hadronic fraction
in the mix phase are independent of numerical values of $n$. However, the behavior of the deceleration parameter $q$ and the dimensionless ratio $\xi/s$ are dependent on $n$. We saw that in QGP phase the temperature of HL early universe filled with causal, full steady, IS fluid, declines faster than the Eckart fluid and is indistinguishable from that of the hadronic fraction $h$ and temperature $T$ in the hadron phase for certain values of $\lambda$.
Also we observed that, in QGP phase, the evolution of the scale factor $a$ for standard cosmology ($\lambda=1$) in both fluid are in agreement.

For non-detailed balance HL world filled with a causal, steady state, full version of the IS relativistic viscous fluid in the QGP phase for $0\leq n \leq0.5$ and $0.5<n\leq1$, we found that $q>0$ and $q<0$ respectively.  In the hadronic phase, for $0\leq n \leq 0.5 $ and
$q>0$,  the rate of the negative acceleration was found to be slowly changing. For $n>1$ the sign of the
deceleration parameter is negative. In both phases, only in the interval $0\leq n\leq 0.8$ for dimensionless ratio $\xi/s$, we observed a compatible behavior with lattice QCD results.

Finally, we  note that in \cite{T.H} the authors have studied the quark-hadron phase transitions (in the absence of viscosity) in the context of alternative models of gravity. In \cite{A.H} however, using a phenomenological model and lattice QCD simulations the authors have analyzed the effects of viscosity on the phase transition in the context of standard cosmology. In the present work, using different phenomenological models, we have studied the effects of the viscosity on the phase
transition in the context of HL early universe. It is worth noting that we have offered two critical temperatures $T_{c}\approx120MeV$ and $T_{c}\approx340MeV$
for first order phase transition, which according to results presented in \cite{A.H}, $T_{c}\approx340MeV$ is unacceptable. Also, as shown in
\cite{A.H} we see that the general behavior of some of the cosmological parameters of the early universe such as temperature $T$ and the scale factor in HL
early universe for $\lambda=1$ is similar to the standard cosmology. The authors in \cite{A.H} have  considered the evolution of the Hubble parameter $H$ in the framework of the Eckart
and full causal IS fluids. However what we have done in this work has been to study other cosmological parameters in the Eckart and steady full causal IS fluids.

\end{document}